\documentclass[bibyear]{aa}

\usepackage{graphicx}
\usepackage{txfonts}
\usepackage{float}
\usepackage{xcolor}

\usepackage{natbib,twoopt}
\usepackage[breaklinks=true,colorlinks=true,linkcolor=blue,citecolor=blue,urlcolor=blue]{hyperref}
\bibpunct{(}{)}{;}{a}{}{,}            
\usepackage{color}
\usepackage[flushleft]{threeparttable}

\begin{document}

\title{Black Holes in the Shadow: The Missing High-Ionization Lines in the Earliest JWST AGNs}

\author{
  Greta Zucchi\inst{1,2}
  \and
  Xihan Ji \inst{3,5}
  \and 
  Piero Madau\inst{4,1}
  \and
  Roberto Maiolino \inst{3,5,6}
  \and
  Ignas Juod\v{z}balis \inst{3,5}
  \and
  Francesco D’Eugenio \inst{3,5}
  \and 
  Sophia Geris \inst{3,5}
  \and 
  Yuki Isobe \inst{3,5,7}
}

\institute{
 Dipartimento di Fisica "G. Occhialini", Università degli Studi di Milano-Bicocca, Piazza della Scienza 3, I-20126 Milano, Italy
  \and
  Stockholm University, Department of Astronomy and Oskar Klein Centre for Cosmoparticle Physics, AlbaNova University Centre, SE-10691, Stockholm, Sweden \\
\email{greta.zucchi@astro.su.se}  
  \and
   Kavli Institute for Cosmology, University of Cambridge, Madingley Road, Cambridge CB3 0HA, UK
   \and
  Department of Astronomy \& Astrophysics, University of California Santa Cruz, 1156 High Street, Santa Cruz, CA 95064, USA
  \and 
  Cavendish Laboratory, University of Cambridge, 19 JJ Thomson Avenue, Cambridge CB3 0HE, UK
  \and
  Department of Physics and Astronomy, University College London, Gower Street, London WC1E 6BT, UK
  \and
  Waseda Research Institute for Science and Engineering, Faculty of Science and Engineering, Waseda University, 3-4-1, Okubo, Shinjuku, Tokyo 169-8555, Japan
}

%\date{Received XXX, YYYY; accepted XXX, YYYY}

\abstract{
Observations with the James Webb Space Telescope (JWST) have uncovered a substantial population of high-redshift, broad-line active galactic nuclei (AGNs), whose properties challenge standard models of black hole growth and AGN emission. We analyze a spectroscopic sample of 34 Type 1 AGNs from the JWST Advanced Deep Survey (JADES) survey, spanning redshifts $1.7 < z < 9$, to constrain the physical nature of the accretion flows powering these sources with broad-line diagnostics statistically for the first time. 
At $z > 5$, we find a marked suppression of high-ionization emission lines (HeII, CIV, NV) relative to prominent broad H$\alpha$ and narrow [OIII] features. This contrast places strong constraints on the shape of the ionizing spectral energy distribution (SED) and on the physical conditions in the broad-line region (BLR). By comparing the observations to photoionization models based on SEDs of black holes accreting at sub-Eddington ratios, we show that standard AGN continua struggle to reproduce the observed broad line ratios and equivalent widths across a wide ionization parameter range. These results suggest the need for modified SEDs -- 
either intrinsically softened due to super-Eddington accretion or radiative inefficiencies in the innermost disk, or externally filtered by intervening optically thick gas that absorbs or scatters the highest-energy photons before they reach the BLR.  
}


\maketitle

\section{Introduction}

Deep-field observations with the James Webb Space Telescope (JWST) have identified a substantial population of moderate-luminosity ($L_{\rm bol}\sim10^{43}$\,-\,$10^{45}~{\rm erg~s^{-1}}$), broad-line active galactic nuclei (AGNs) at redshifts $z > 4$ [with the full width at half maximum (FWHM) of broad Balmer lines usually being 1000\,-\,6000 $\mathrm{km \, s^{-1}}$], with number densities that are nearly 100 times higher than those of UV-selected quasars at similar epochs \citep[e.g.,][]{Harikane2023AGN,juodvzbalis2025a}. These AGNs are powered by accretion onto early massive black holes (MBHs) that appear overmassive with respect to the local black hole mass -- stellar mass relation (e.g., \citealp{cohn_2025,Harikane2023AGN,Kocevski2023}, but see also \citealp{geris2025}). They also show unusually weak X-ray emission \citep[e.g.,][]{Ananna_lrd_2024,Kokubo_2024,maiolino2024a,Yue2024, Sacchi2025}, faint high-ionization lines \citep[][]{Lambrides2024,juodvzbalis2025a,wang_heii_2025} with only a few outliers reported by \citet{Tang2025}, and in some cases Balmer line absorption features \citep[especially for those classified as ``Little Red Dots'', e.g.,][]{Matthee2024,juodzbalis_2024,Kocevski2025,ji_qso1_2025,deugenio_restabs_2025,deugenio_qso1_2025} that are suggestive of large column densities of neutral hydrogen in the nuclear region \citep[e.g.,][]{juodzbalis_2024,im25,ji_qso1_2025,degraaff2025,naidu_2025,taylor_lrd_2025,deugenio_irony_2025}. These properties present substantial challenges to current theories of MBH growth and the nature of broad-line AGNs at high redshift.

Rest-frame UV-optical emission lines in AGN spectra are highly sensitive diagnostics of the shape of the SED, particularly the part of the SED in the far-ultraviolet (FUV), extreme ultraviolet (EUV), and soft X-ray bands -- wavelength ranges that are largely inaccessible due to interstellar and intergalactic absorption \citep[e.g.,][]{Osterbrock2006}. Variations in line ratios and equivalent widths (EWs) encode information about the physical conditions in the line-emitting gas, including ionization parameter, metallicity, covering factor, geometry, and -- crucially -- the Eddington ratio $\lambda_{\rm Edd} = \dot{M}/\dot{M}_{\rm Edd}$ and black hole mass \citep{BorosonGreen1992, Ferland2020}. The degeneracies inherent in this mapping have historically limited the diagnostic power of photoionization models, particularly due to the poorly constrained shape of the ionizing continuum between the Lyman limit and \(\sim 0.3\,\mathrm{keV}\) \citep[e.g.,][]{Done2012}.

Recent advances in SED templates for sub-Eddington and super-Eddington accretion flows have enabled a more quantitative comparison between predicted and observed emission line trends \citep[e.g.,][]{Capellupo2015, Hall2018, Kubota2019}. Some of these models also incorporate geometric effects and the anisotropic radiation fields inherent to supercritical accretion flows \citep[e.g.,][]{Wang2014, Pacucci2024, Madau2024, Madau2025}. Although bolometric luminosities and black hole masses inferred from deep JWST surveys typically suggest sub-Eddington accretion rates for the general population of AGNs at $z\gtrsim4$ \citep[e.g.,][]{maiolino2024b,juodvzbalis2025a}, large uncertainties in virial mass estimates as well as bolometric conversions, particularly at high redshift, leave the Eddington ratio poorly constrained \citep[e.g.,][]{Lupi2024,Bertemes2025}. There are only a few cases in which the black hole masses have been measured not through the single-epoch method, but more directly through dynamics in $z>2$ AGNs, revealing both super-Eddington accretions \citep[in a few luminous quasars, e.g.,][]{abuter2024}, and sub-Eddington accretions \citep[in the case of a lensed LRD at $z=7$,][]{juodvzbalis2025b}.

The anomalies observed in JWST AGNs raise key questions about the shape of the ionizing continuum, the structure and physical conditions of the BLR, and the nature of radiative transfer in early AGNs. The marked deficiency of high-ionization emission lines suggests a significant departure from the standard AGN photoionization paradigm. Two primary scenarios may account for this: either the ionizing SEDs are intrinsically soft or filtered, reducing the photon flux at energies $\gtrsim 50$ eV; or the AGNs are accreting in a super-Eddington regime, where geometric effects, inner obscuration, and radiation collimation lead to anisotropic emission and selective suppression of high-ionization lines \citep{Madau2025}. These possibilities carry important implications for the physical state of the BLR and the fueling mechanisms of early black holes.

To investigate the physical origin of the emerging spectroscopic trends in early AGNs, we analyze a sample of 34 broad-line AGNs from the JWST JADES survey \citep{Bunker_dr1,Rieke_dr1,eisenstein2023overview,DEugenio2024} selected by \citet{juodvzbalis2025a}, spanning the redshift range $1.7 < z < 9$. Using both prism and grating spectroscopy, we examine rest-frame UV and optical emission-line diagnostics to constrain the ionization structure and continuum hardness across cosmic time. We construct a grid of photoionization models varying in ionization parameter, gas density, and incident SED shape, and compare predicted line ratios and equivalent widths to the observations. Special emphasis is placed on the $z > 5$ regime, where deviations from local AGN templates seem  most pronounced and where constraints on the ionizing continuum are most urgently needed. In this work, we focus on the regime of sub-Eddington accretion and assess whether standard SEDs representative of nearby AGNs can reproduce the observed line properties. The limitations of these models provide a benchmark for future investigations of super-Eddington accretion and SED filtering scenarios, which will be explored in a forthcoming paper.

\section{Sample Selection}
\label{sec:data_sel}

We construct our sample from the Type 1 AGNs selected by \citet{juodvzbalis2025a} who used JWST NIRSpec medium resolution spectra ($R \sim 1000$) for identifications of BLR emission. 
Emission lines were fit using multiple Gaussian components to distinguish narrow-line emission from star formation or the narrow-line region, and broad emission from high-velocity virialized gas near the black hole.
H$\alpha$ was used as the primary tracer due to its high luminosity and visibility in JWST spectra over $z \sim 0.5$-7.

In the following, we summarize the identification and selection criteria as performed by \citet{juodvzbalis2025a}.
In their analysis, for each source, the H$\alpha$+[NII] emission complex was fitted with two models: one model with only narrow H$\alpha$ and [NII]$\lambda\lambda$6548,6583 emission (constrained to share velocity and width, with the [NII] doublet ratio fixed to 3), and the other model with an additional broad H$\alpha$ component. Markov Chain Monte Carlo (MCMC) sampling was used to fit parameters. 
The same method was applied to [OIII]$\lambda\lambda$4959,5007 and H$\beta$ to identify any ionized outflows and estimate dust attenuation via the Balmer decrement.
In cases with significant outflow features, H$\alpha$ was refit including an outflow component with priors derived from [OIII], to distinguish outflows from the BLR. Since BLRs do not have [OIII] emission, excess broad wings in H$\alpha$ lacking [OIII] support were attributed to the BLR. 
Within the JADES sample, four sources require further analyses as noted by \citet{juodvzbalis2025a}.
\begin{itemize}
    \item GN-200679 (z = 4.547) lacked [OIII] coverage in its NIRSpec spectra, so the broad H$\alpha$ identified may also trace an outflow -- thus flagged tentative;
    \item GN-23924 (z = 1.676) showed a weak BLR compared to outflows, also marked tentative;
    \item GS-49729 (z = 3.189) and GS-159717 (z = 5.077) exhibited complex broad-line profiles only well fitted with two Gaussians or an exponential model, with GS-159717 further requiring a strong rest-frame absorption component \citep{juodvzbalis2025a,deugenio_restabs_2025}. The complex profile potentially reflecting non-Gaussian BLR kinematics \citep{kollatschny2013}, unresolved black hole binaries \citep{maiolino2024b}, or radiative transfer effects (\citealp{Laor_2006,rusakov_2025}, see, however, \citealp{brazzini_2025} for a counter argument). For consistency, the two broad components were combined into a single one to estimate FWHM and luminosity.
\end{itemize}
Using the methodology described above, a total of 30 Type 1 AGNs were identified by \citet{juodvzbalis2025a}, with 28 robust identifications based on the presence of a broad H$\alpha$ component with signal-to-noise (S/N) > 5, and 2 tentative identifications that narrowly miss one of the selection thresholds. To compare the goodness-of-fit of models with and without a broad component, indeed, they used also the Bayesian Information Criterion (BIC) with $\Delta \mathrm{BIC} > 5$, which penalizes model complexity and it is defined as $\mathrm{BIC} = \chi^2 + k \ln n$, with $k$ the number of free parameters and $n$ the number of data points.  

Visual inspections of the low-resolution NIRSpec/prism spectra ($R\sim 100$) led to the identification of four additional Type 1 AGNs at $7 < z < 9$ (GN-4685, GS-20030333, GS-20057765, GS-164055) showing hints of broad H$\beta$ emission. While individually these sources did not meet formal significance thresholds, their stacked spectrum, weighted by inverse variance, revealed a statistically significant broad H$\beta$ component. This detection was confirmed via jackknife resampling, and we include them here following \citet{juodvzbalis2025a} for completeness. Therefore, these four objects were added to the Type 1 sample, raising the total AGN number to 34 spanning redshift $1.7 < z < 9$.
The sample is twice as large as that used in the JWST AGN SED study by \citet{Lambrides2024}, and spans a nominal black hole mass range of $M_{\rm BH} \sim 10^{6}$–$10^{9}~M_\odot$ and Eddington ratios from 0.02 to 0.4, 
derived from single-epoch virial black hole mass estimates and bolometric luminosities, based on H$\alpha$ as described by \citet{juodvzbalis2025a} following the relation provided by \citet{reines2015}. However, as we will show later, despite the moderate nominal $\lambda _{\rm Edd}$ for the bulk of our sample, standard sub-Eddington SEDs do not provide a satisfactory fit to the observed emission.

Notably, all of these AGNs, except GS-49729 and GS-209777, are not detected in X-ray at roughly 2\,-\,10 keV in the rest frame, implying significant X-ray weakness compared to local AGNs \citep{maiolino2024a,juodvzbalis2025a}.
As we will show later, the two X-ray detected sources also show drastically different broad-line spectra compared to other sources, and they are more similar to local sources.
Next, we describe the spectral measurements we made to our sample.

\section{Spectral Measurements}
\label{sec:stacked_spectra}

\subsection{Stacking}

To better constrain intrinsically weak or low S/N spectral features, we performed spectral stacking across selected redshift intervals. We applied this method to both the low-resolution ($R\sim 100$) NIRSpec prism spectra and the medium-resolution ($R\sim 1000$) grating spectra within our sample, using the latest data products reduced with the NIRSpec GTO pipeline \citep[see e.g.,][]{Carniani_2024,DEugenio2024}, which incorporates the most recent JWST calibration updates.

As we mentioned, two JADES AGNs (IDs GS-49729 and GS-209777) have X-ray detections and we excluded them from the stack.
Interestingly, they are more similar to X-ray normal AGNs/QSOs in terms of other spectral features as well, with GS-49729 showing an extremely strong CIV line not visible in the rest of the sample, and both sources displaying QSO-like continua \citep{juodvzbalis2025a}.
Their continuum shape would bias the stacked spectrum relative to the X-ray undetected population studied here. In addition, GS-209777 shows a steeper Balmer decrement in the narrow lines compared to other objects \citep[see Table 2 in ][]{juodvzbalis2025a}, suggesting that its spectral properties differ from those of the bulk of the sample.

\begin{figure}
    \centering
    \includegraphics[width=0.9\columnwidth]{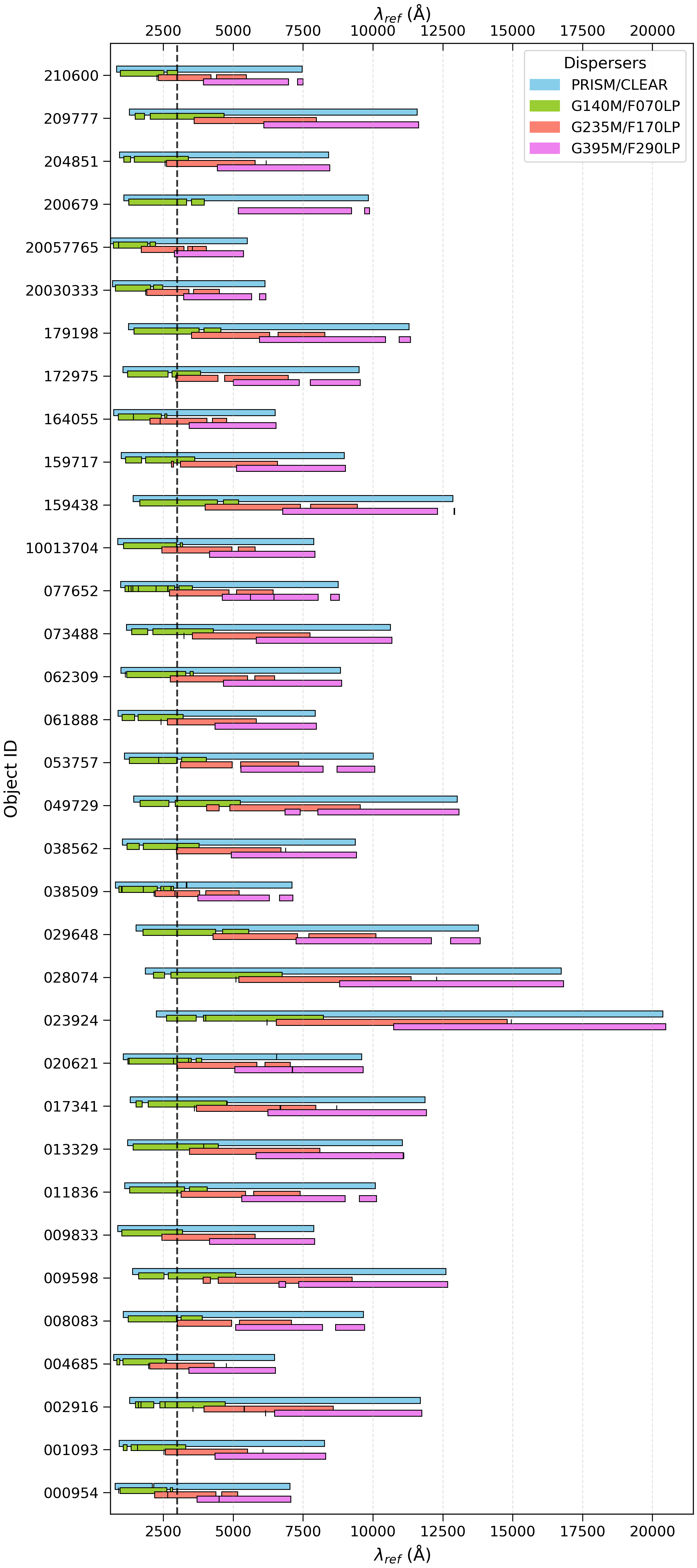}
    \caption{Spectral wavelength coverage across dispersers for each object. Invalid or missing values are masked and shown as empty. The dashed vertical line at 3000 \AA\ indicates a conventional division between the UV and optical regimes, though it is not used in the kinematic classification of emission lines. Treatment of lines near this boundary (e.g., [NeV]$\lambda$3426) is discussed in Section~\ref{sec:spectralfitting}.
    }
\label{fig:disp_wl_cover}
\end{figure}

For the remaining X-ray undetected sources, each prism/grating spectrum was first shifted to the rest frame using the spectroscopic redshift. The spectra were then resampled onto a common wavelength grid with a 2\,\AA\ step for grating and a 5\,\AA\ step for prism using \textsc{SpectRes} \citep{carnall2017spectres}. The chosen wavelength steps ensure consistent sampling across all spectra, enabling accurate alignment of spectral features and maximizing sensitivity to faint emission lines: using much finer sampling would only amplify noise fluctuations without adding information, whereas coarser sampling would risk smoothing out the spectral features.
The stacking wavelength range in each redshift bin was chosen to ensure at least 70\% spectral coverage throughout the sample. We found that the prism provides rest-frame UV coverage for all 32 objects, while among the gratings only G140M sufficiently covers the UV, with 25 out of 32 spectra contributing. In the optical regime, no single grating spans the full range; instead, it is covered in pieces by G140M, G235M, and G395M. Therefore, all measurements in the optical range are based on stacks that combine the spectra from these three gratings. Figure \ref{fig:disp_wl_cover} shows the rest-frame wavelength coverage of each disperser for all individual targets.

Stacking was performed as an inverse-variance weighted average to maximize S/N, normalized by the flux F[OIII]$\lambda$5007, following \citet{isobe2025jades}. For grating spectra, line fluxes were measured by fitting Gaussian profiles, with uncertainties estimated via Monte Carlo resampling. For prism spectra, where the [OIII] doublet is blended and the low resolution provides too few points for a reliable Gaussian fit, we measured the flux by direct summation of the continuum-subtracted line profile, and derived the [OIII]$\lambda5007$ flux by scaling from the total doublet flux; the uncertainty was obtained by summing the per-pixel errors in quadrature over the same interval.
The [OIII] flux measured from the prism spectrum was used for normalization when no grating-based measurement was available for a given object. These measurements were used solely for normalization purposes prior to stacking, while the detailed spectral fitting of the stacked spectra is described in Section~\ref{sec:spectralfitting}.
Normalized spectra and their associated normalized uncertainties were combined on a common grid removing outlier pixels via 5$\sigma$ sigma-clipping. Finally, the stacks and their formal errors were obtained as the inverse-variance weighted mean and the corresponding weighted standard error of the mean, respectively.

During stacking, we divide our sample into three redshift bins ($z < 3.5$, $3.5 < z < 5$, and $z > 5$).
Since, as described above, for each redshift bin, we stacked the prism, G140M grating, and combined gratings for different spectral ranges, we obtained a total of $3\times 3=9$ stacked spectra, three at $R\sim 100$ and six at $R\sim 1000$. Figure \ref{fig:z5_grating_stack} shows one of the resulting stacked spectra, the others are reported in the Appendix \ref{appendix}.

\begin{figure*}
    \centering
    \includegraphics[width=\textwidth]{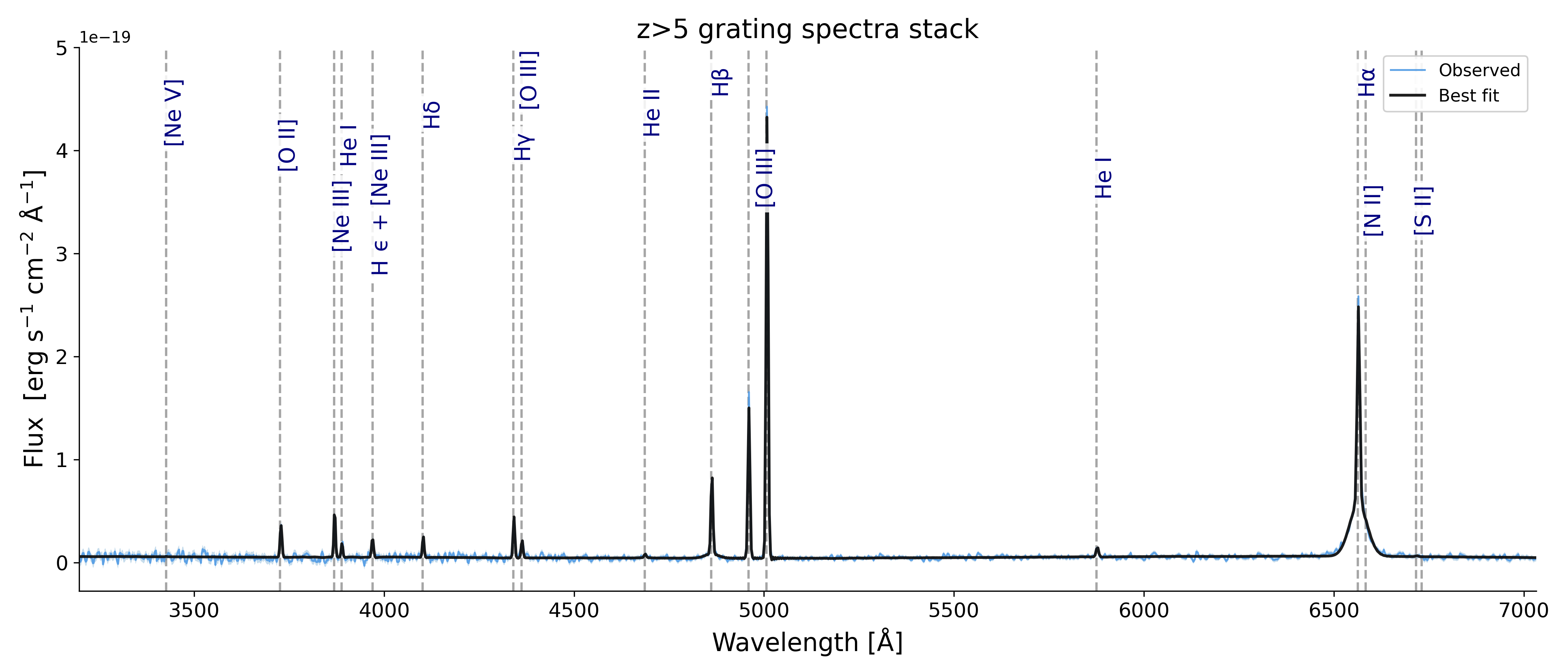}
    \caption{Stack of the JWST NIRSpec (G140M+G235M+G395M) grating spectra of JADES Type 1 AGNs in the high-redshift bin ($z>5$). For visualization, the normalized stack was multiplied by the mean [OIII]$\lambda$5007 flux of the 15 contributing objects; this rescaling is used only for the figure and not for any measurements.}
    \label{fig:z5_grating_stack}
\end{figure*}

\subsection{Spectral Fitting}
\label{sec:spectralfitting}

Building on the resolution-matched stacked spectra, we initiated the spectral fitting process of the stacked spectra to recover emission-line fluxes and kinematics, accounting for instrumental broadening via adopting the MSA line spread function (LSF) for point-like sources from \citet{degraff2024}, evaluated at the average redshift of the sources in each stack. 
In the case of mixed grating stacks (G140M+G235M+G395M), where there are three different LSFs, we adopted the LSF of the grating that contributes most to each redshift bin (i.e., G235M as it covers the majority of objects).

We then performed a full-spectrum fitting using a modified version of the Python implementation of the Penalized PiXel-Fitting \citep[pPXF;][]{cappellari2004, cappellari2017, cappellari2023} method customized for AGN analysis. The observed spectra were modeled as a combination of stellar continuum templates and emission-line components, with the AGN continuum approximated by high-order Legendre additive polynomials rather than an explicit template \citep[e.g.,][]{marconcini2025, degraff2025}, to optimize the fit of the continua and nebular emission simultaneously. 
For the stellar continuum, we used templates from the eMILES stellar population synthesis library \citep{vazdekis2016}, 
which encompasses a broad range of stellar ages and metallicities, ensuring a realistic representation of stellar absorption features and continuum shapes in the wavelength range 1680 -- 50,000 \AA. We modeled emission-line profiles as Gaussians, including narrow emission lines and broader components associated with the BLR. For emission-line doublets that arise from the same upper energy level, the flux ratios are governed by atomic physics and held fixed\footnote{The fixed ratios were applied to the [OIII]$\lambda\lambda$4959,5007, 
[NII]$\lambda\lambda$6548,6583, 
[NeV]$\lambda\lambda$3346,3426, 
[NeIII]$\lambda\lambda$3869,3968, 
CIV$\lambda\lambda$1548,1551, 
NV$\lambda\lambda$1238,1242, 
and MgII$\lambda\lambda$2796,2803 doublets, with the fluxes of the fainter lines scaled from the brightest component in each pair.}. We adopted the theoretical values computed with the \texttt{PyNeb} library \citep{luridiana2015pyneb}, using them as fixed scaling factors during both the construction of the line templates and the measurement of integrated fluxes.\\

To evaluate emission-line detections, we applied a conservative $3\sigma$ threshold based on the formal uncertainties returned by \texttt{pPXF}. A line was considered detected if its integrated flux exceeded three times its uncertainty; otherwise, we report a $3\sigma$ upper limit. Integrated fluxes were computed by multiplying the best-fit amplitude of each line by the local spectral dispersion and the continuum normalization factor.

\subsubsection{The G140M+G235M+G395M stack}
The combined-grating stacked spectra -- used for the measurements of optical emission lines -- were modeled using additive and multiplicative Legendre polynomials. For the $3.5 < z < 5$ and $z > 5$ stacks, we adopted a $6^{\mathrm{th}}$-degree additive polynomial together with a $2^{\mathrm{nd}}$-degree multiplicative polynomial. Since our focus is on emission lines, the continuum modeling serves only as a baseline and is not interpreted physically.
For the $z < 3.5$ stack, however, a higher degree of flexibility was required, and we used a $16^{\mathrm{th}}$-degree additive polynomial and a $4^{\mathrm{th}}$-degree multiplicative polynomial. This likely reflects the limited number of spectra contributing to this stack (only five objects), which made it more challenging to constrain the continuum shape. We also allowed for line asymmetry -- broad emission lines were modeled with Gauss–Hermite profiles including the $h_3$ and $h_4$ terms, enabling deviations from purely Gaussian symmetric shapes, without attributing them to specific kinematic sub-components such as outflows. The low-redshift stack demanded this additional freedom, likely due to the smaller sample size and the correspondingly more complex continuum and line profiles.
At this spectral resolution, we created three distinct kinematics groups: 

\begin{enumerate}

    \item {\bf Narrow components}: H$\alpha$, H$\beta$, H$\epsilon$, H$\delta$, H$\gamma$, [OII]$\lambda\lambda$3726,3729,  HeI$\lambda$3889, [OIII]$\lambda$4363, HeII$\lambda$4686, HeI$\lambda$5876, [NII]$\lambda\lambda$6548,6583, [SII]$\lambda\lambda$6716,6731;\\
    
    \item {\bf High-ionization narrow components}: [NeV]$\lambda$3426, [NeIII]$\lambda\lambda$3869,3968, [OIII]$\lambda\lambda$4959,5007; and \\
    
    \item {\bf Broad components}: H$\alpha$, H$\beta$, HeII$\lambda$4686.
\end{enumerate}

\noindent High-ionization narrow lines likely originate closer to the AGN and are more susceptible to the kinematic effects of AGN-driven outflows or turbulent motions within the narrow-line region (NLR) and are in general found to be broader than the other narrow lines \citep[e.g.,][]{spoon2009, dasyra2011}. Both a narrow and a broad HeII$\lambda$4686 component were included in the fit to ensure proper constraints on the BLR emission, as both narrow and broad HeII have been observed in JWST/NIRSpec studies of high-redshift AGNs \citep[e.g.,][]{ubler2023}, and the presence of a broad HeII feature is commonly observed in luminous Type 1 AGNs \citep[e.g.,][]{marziani2010}. 
In the meanwhile, the narrow HeII has been proved to be a key diagnostic of EUV photon sources in galaxies and shown to be effective at distinguishing AGN from stellar ionization in low-metallicity environments \citep[e.g.,][]{shirazi2012,katz_2023,ubler2023,Cleri_2025,Scholtz_2025,treiber_2025}.

Across both prism and grating stacked spectra, broad components were only fitted for a restricted set of lines. Although in principle all permitted lines may display a broad component, in practice we limited our analysis to H$\alpha$, H$\beta$, CIV, and HeII. Broad components in other transitions are generally too weak or poorly constrained in our data to yield reliable results, whereas CIV and HeII were nonetheless included since establishing constraints (including upper limits) on these lines is a key goal of this work.\\

\subsubsection{The prism stack}
For the prism stacked spectra, we adopted a slightly modified fitting strategy. 
In this case, we employed a $12^{\mathrm{th}}$-degree additive polynomial, allowing for line asymmetries to ensure sufficient flexibility. 
We also tested fits with different levels of high-order polynomials and found that the resulting line fluxes remained stable to within $\sim 10\%$.

The broader wavelength coverage of the prism spectra enabled us to fit UV emission lines, which could not be consistently accessed with the gratings except for the G140M stack -- with the specific procedures for tying and treating these lines described later in this section. We restricted the fitted range of prism stack spectra to $\lambda _{\rm rest} >1250$\,\AA, thereby excluding the region affected by Ly$\alpha$ and potential damping wings. This conservative cut is motivated by the complex radiative transfer of Ly$\alpha$ photons, which are subject to resonant scattering and significant absorption by the intergalactic medium (IGM).
We further divided the sample into three distinct kinematic groups:

\begin{enumerate}
    \item \textbf{Narrow components:} H$\alpha$, H$\beta$, H$\epsilon$, H$\delta$, H$\gamma$,[OII]$\lambda\lambda$3726,3729, [NeIII]$\lambda\lambda$3869,3968, HeI$\lambda$3889, [OIII]$\lambda$4363, HeII$\lambda$4686, HeI$\lambda$5876, [NII]$\lambda\lambda$6548,6583, [SII]$\lambda\lambda$6716,6731;\\
    
    \item \textbf{UV narrow components:} SiII$\lambda$1260, SiII$\lambda$1304, CIV$\lambda$1549, HeII$\lambda$1640, SiIII]$\lambda$1892, [NeIV]$\lambda$2424, OII]$\lambda$2471, NV$\lambda$1240, OIII]$\lambda\lambda$1661,1666, CIII]$\lambda\lambda$1907,1909, MgII$\lambda$2796, [NeV]$\lambda$3426; and \\
    
    \item \textbf{Broad components:} H$\alpha$, H$\beta$, CIV$\lambda$1549, HeII$\lambda$1640, HeII$\lambda$4686.
\end{enumerate}

\noindent In this classification, we grouped [NeV]$\lambda$3426 with the UV narrow lines. High-ionization UV transitions such as CIV and NV are established tracers of BLR and AGN-driven outflows, often exhibiting broad, blueshifted profiles indicative of fast, ionized gas. Given its similarly high ionization potential, [NeV]$\lambda$3426 is commonly emitted from the inner NLR or coronal-line region and shows comparable kinematic signatures to these UV lines, supporting a shared origin \citep[e.g., AGN photoionization, outflow, shocks, etc.,][]{Thuan_2005,Izotov_2012,izotov_2021,feltre2016,olivier_2022,cleri_ne53_2023,cleri2023,negus_2023,richardson_2025}.
CIV$\lambda$1549 was included in both the broad and narrow groups to account for its emission from multiple regions: the fast, highly ionized gas near the SMBH and more extended, lower-density outflows \citep[e.g.,][]{coatman2016}.
Similarly, HeII$\lambda$1640 was assigned to both the narrow and broad groups, consistent with our treatment of HeII$\lambda$4686 in the optical. Broad HeII$\lambda$1640 components are commonly observed in AGN samples at z$\sim$2-5, where they trace BLR emission, while narrower counterparts arise from more extended ionized gas \citep[e.g.,][]{nanayakkara2019, saxena2020}.

\subsubsection{The G140M stack}
Finally, G140M stacks were introduced to enable measurements in the UV portion of the spectra, allowing us to constrain emission lines down to NV$\lambda$1240. However, no stack was constructed for the $z < 3.5$ bin, as at least 70\% of the spectral range must lie within the G140M instrumental bandpass for inclusion. As a result, only the $3.5 < z < 5$ and $z > 5$ redshift bins are included in the analysis. For these spectra, we adopted the same kinematic groupings used in the prism stacks, but we excluded the optical emission lines, which fall outside the wavelength coverage of the G140M spectrum. Since the UV spectra are generally more noisy compared to the optical spectra, to make the fit of weak lines more robust, the FWHM of each emission component was fixed to the values obtained from the corresponding optical fits.

To discriminate emission-line detections, unlike in other stack analyses, we complemented the formal \texttt{pPXF} uncertainties with an empirical estimate of the local continuum scatter, in order to account for possible underestimation of the errors when the model does not adequately reproduce the observed line profile. Specifically, for each line, we measured the per-pixel root-mean-square (rms) of the residual noise in nearby, line-free continuum windows, using a sigma-clipped median deviation. Comparing the formal and empirical error estimates revealed cases where the \texttt{pPXF}-based significance was likely overestimated. We therefore retained the \texttt{pPXF} fluxes but adopted the continuum-based signal-to-noise ratio as our detection criterion, requiring a minimum of $3\sigma$ for a line to be considered detected.

\subsection{Equivalent Width Measurements}
\label{sec:ew}

The EW of each line was measured as the ratio between the integrated line flux and the underlying continuum level.

For each line, we defined a spectral window centered on its expected wavelength, with a width adapted to the expected kinematic broadening -- narrower for forbidden lines and broader for broad-line components. Within this window, the local continuum level was evaluated using a scale matched to the spectral resolution: for narrow lines we adopted 20 \,\AA\ windows for the gratings and 80 \,\AA\ for the prism, while for broad lines we used 60 \,\AA\ and 120 \,\AA\, respectively. To improve the reliability of the continuum estimate, we leveraged the higher-S/N continuum from the prism spectra even when analyzing lines measured with the grating \citep[see JADES DR3 paper,][]{DEugenio2024}. This choice is further motivated by spectral overlap in the more dispersed grating data, which can complicate a direct continuum determination from the gratings alone. Specifically, we evaluated the best-fit prism continuum model on the grating wavelength grid via linear interpolation. This prism-derived continuum level was used for all EW calculations -- regardless of the disperser -- while the integrated line fluxes and their uncertainties were taken from either the prism or grating fit, depending on the spectrum.

An exception was made for the G140M stacks. At $z > 5$, the damped Ly$\alpha$ absorber (DLA) is poorly resolved in the prism data, leading to unreliable continuum estimates near certain UV lines, such as NV$\lambda1240$. Although the overall prism and G140M continua are broadly consistent, these localized issues motivated us to rely exclusively on the G140M continuum when computing EWs in the $z > 5$ and $3.5 < z < 5$ bins.

To account for uncertainties in the continuum level and line integration window in the determination of the EW, we adopted a Monte Carlo approach that robustly quantifies EW variations. 
In each MC realization, we introduced small random shifts to both the line center and the integration window, and re-evaluated the continuum level as the median flux within the window on the best-fit continuum model. The EW was then calculated for each trial, and the resulting distribution was used to derive the mean EW value.
To estimate EW uncertainties, thus, we combined three contributions in quadrature: (i) the statistical uncertainty on the line flux returned by the spectral fit; (ii) the scatter in EW from Monte Carlo sampling of the continuum level and line window; and (iii) the uncertainty on the continuum level itself. The latter was estimated from the standard deviation of the residuals in a local window around the line, excluding known emission features to avoid contamination. For lines with both narrow and broad components, fluxes and uncertainties were summed in quadrature before computing the total EW and its associated error.

To estimate the continuum-noise contribution to the EW uncertainty, we used the residuals from the prism continuum fit (except for the G140M stacks). This approach ensures that uncertainties in the continuum normalization are consistently propagated across all spectra, while the line fluxes themselves remain anchored to the grating data. Because broad components are reliably measured only in the grating spectra -- not in the lower-resolution prism data -- we report total (narrow + broad) fluxes in the prism case, but decompose the broad-line contributions where permitted by the grating fits.

Table \ref{tab:ew} reports the measured rest-frame EWs of emission lines from the stacked grating (G) and prism (P) spectra, separated into three redshift bins, distinguishing between narrow and broad components and providing upper limits where lines are undetected. In the $z > 5$ stacks, Balmer lines such as H$\beta$ and H$\alpha$ are clearly detected and remain strong. For example, the H$\beta$ EW is measured at $\sim$37\,\AA\ (broad, grating) and $\sim$115\,\AA\ (total, prism), while H$\alpha$ (broad) reaches $\sim$540–700\,\AA, indicating a substantial covering factor and efficient reprocessing of ionizing radiation in the BLR. The helium recombination line HeI$\lambda$5876 is also robustly detected with an EW of $\sim$16–22\,\AA, making it another tracer of low-ionization conditions.

In contrast, high-ionization lines such as CIV$\lambda$1549 and HeII\,$\lambda$1640 remain undetected or weak in the $z > 5$ stacks, both for the grating and prism data. Upper limits for HeII$\lambda$4686 (broad) are $\lesssim$13–35\,\AA, and no significant detections are reported for CIV -- a line that is nearly ubiquitous in lower-redshift Type 1 AGNs \citep[e.g.,][]{vandenberk_2001,richards_2011,Wu2022}. This selective suppression of high-ionization lines suggests a low ionization parameter, a soft or filtered AGN SED, or possibly super-Eddington accretion scenarios in which the ionizing continuum is anisotropic or collimated away from the BLR.

The forbidden narrow lines further constrain the shape and reach of the ionizing spectrum. [NeIV]\,$\lambda$2424 (ionization potential $\sim$63\,eV) is undetected in the grating stacks but appears weakly in prism data at $z > 5$ (EW $\lesssim$4.2\,\AA), while [NeV]\,$\lambda$3426 (ionization potential $\sim$97\,eV), a classical tracer of very hard continua, is absent above $z>3.5$. The [OIII]\,$\lambda$5007 line, by contrast, is detected at all redshifts, with equivalent widths of $\sim$570–600\,\AA\ even at $z > 5$. Although [OIII] is not a very high-ionization tracer -- its ionization potential is $\sim$35\,eV, lower than 54 eV that is usually used for defining very high-ionization lines \citep[e.g.,][]{berg_2021,olivier_2022} -- its persistence indicates a significant moderate-energy photon flux, even when the far-UV or soft X-rays appear suppressed.

\begin{figure}
    \centering
    \includegraphics[width=\columnwidth]{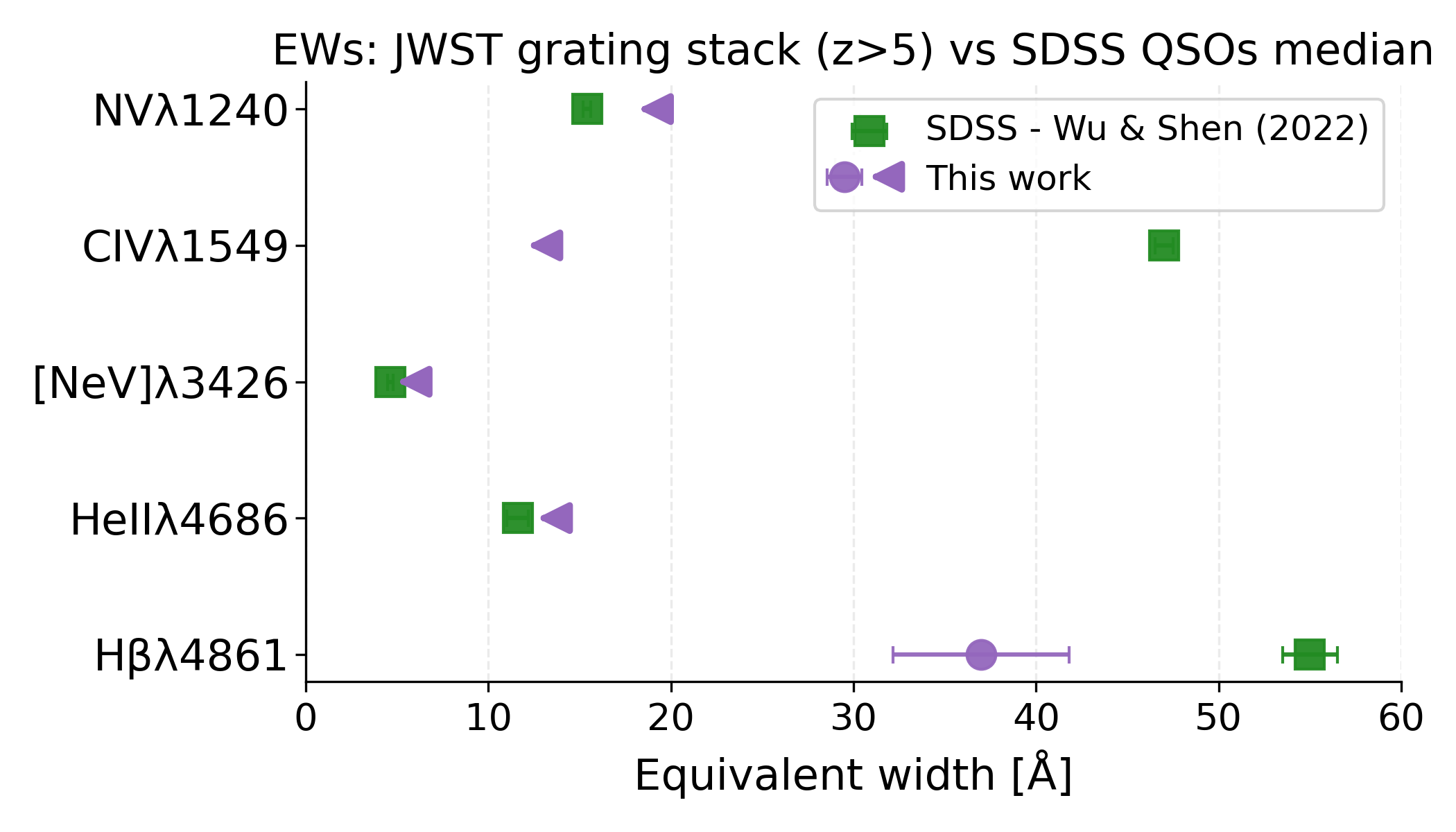}
    \caption{Comparison of emission-line EWs in the JWST grating stack for the $z > 5$ sample (purple) with median values from the SDSS quasar catalog of \citet{Wu2022} (green). Circles and squares represent measured lines with $1\sigma$ uncertainties, while left-pointing arrows denote $3\sigma$ upper limits for undetected lines in the JWST stack. Only the broad components are considered for C\,IV, He\,II, and H$\beta$. In most  cases, error bars are smaller than the marker symbols and are not visually apparent.}
    \label{fig:ew_jwst_sdss}
\end{figure}

For comparison with our high-redshift composite, we adopt the median EWs from the SDSS DR16 broad-line quasar sample analyzed by \citet{Wu2022}. Their study, based on over $7.5 \times 10^5$ quasars spanning $0.1 < z < 6$ and bolometric luminosities in the range $44 \lesssim \log(L_{\mathrm{bol}}/{\rm erg\,s^{-1}}) \lesssim 48$, provides robust emission-line measurements for luminous AGNs across cosmic time. As shown in Figure~\ref{fig:ew_jwst_sdss}, the EW of broad H$\beta$ in the $z > 5$ stack ($37 \pm 5$\,\AA) is significantly lower than the SDSS median of $55 \pm 1.5$\,\AA. Even more striking is the non-detection of CIV, with an upper limit of ${\rm EW}<13$\,\AA\ compared to a typical SDSS value of 47\,\AA, indicating a marked suppression of high-ionization UV lines. Upper limits for NV, HeII, and [NeV] are comparable to or only slightly above their SDSS counterparts, suggesting that these lines are at most as strong as in local quasars.

These results highlight a systematic difference in the ionizing spectra and BLR conditions of high-redshift AGNs compared to their lower-redshift analogs. The weakness or absence of high-ionization lines (CIV, HeII, [NeV]) -- relative to the robust Balmer and [OIII] emission -- places strong constraints on the shape of the ionizing SED and the BLR geometry. As we discuss below, such trends may favor either (1) soft, filtered continua due to radiative transfer effects or shielding in the inner disk, or (2) super-Eddington accretion scenarios in which the ionizing output is intrinsically anisotropic and collimated, illuminating only portions of the BLR.

\begin{table*} %test table
\centering
\caption{\footnotesize Rest-frame equivalent widths (in \AA)  of prominent emission lines measured in stacked JWST/NIRSpec spectra of Type 1 AGNs, grouped by redshift and spectroscopic mode (Grating = G, Prism = P); values are given separately for broad and narrow components where applicable. Note: lines limits are given in narrow components unless otherwise specified.}

\begin{threeparttable}
\label{tab:ew}
{\scriptsize
\begin{tabular}{ccccc}
\hline
\noalign{\smallskip}
Line & Stack & EW($z < 3.5$) & EW($3.5 < z < 5$) & EW($z > 5$) \\
\noalign{\smallskip}
\hline
\noalign{\smallskip}
$\mathrm{N\,V}\lambda1240$ & G & $-$\tnote{\ddag} & $-$\tnote{\ddag} & $< 19.09$ \\
$\mathrm{Si\,II}\lambda1260$ & G & $-$\tnote{\ddag} & $-$\tnote{\ddag} & $< 18.39$ \\
$\mathrm{Si\,II}\lambda1304$ & G & $-$\tnote{\ddag} & $-$\tnote{\ddag} & $< 11.16$ \\
$\mathrm{C\,IV}\lambda1549_\mathrm{narrow}$ & G & $-$\tnote{\ddag} & $< 13.56$ & $< 13.04$ \\
$\mathrm{C\,IV}\lambda1549_\mathrm{broad}$  & G & $-$\tnote{\ddag} & $< 13.81$ & $< 13.03$ \\
$\mathrm{C\,IV}\lambda1549_\mathrm{total}$  & G & $-$\tnote{\ddag} & $< 27.37$ & $< 26.07$ \\
$\mathrm{He\,II}\lambda1640_\mathrm{narrow}$ & G & $-$\tnote{\ddag} & $< 6.81$ & $< 17.80$ \\
$\mathrm{He\,II}\lambda1640_\mathrm{broad}$  & G & $-$\tnote{\ddag} & $< 7.72$ & $< 17.78$ \\
$\mathrm{He\,II}\lambda1640_\mathrm{total}$  & G & $-$\tnote{\ddag} & $< 14.53$ & $< 35.57$ \\
$[\mathrm{O\,III}]\lambda1661$ & G & $-$\tnote{\ddag} &$< 7.70$  & $< 19.40$ \\
$[\mathrm{O\,III}]\lambda1666$ & G & $-$\tnote{\ddag} &$< 6.98$  & $< 19.78$ \\
$\mathrm{Si\,III}]\lambda1892$ & G & $-$\tnote{\ddag} & $< 5.14$ & $< 10.08$ \\
$\mathrm{C\,III}]\lambda1907$ & G & $-$\tnote{\ddag} & $5.69 \pm 0.87$  & $< 9.90$ \\
$\mathrm{C\,III}]\lambda1909$ & G & $-$\tnote{\ddag} & $< 5.25$ & $< 10.03$ \\
$[\mathrm{Ne\,IV}]\lambda2424$ & G & $-$\tnote{\ddag} & $< 5.65$ & $< 55.39$ \\
$\mathrm{O\,II}]\lambda2471$ & G & $-$\tnote{\ddag} & $< 5.76$ & $< 47.04$ \\
$\mathrm{Mg\,II}\lambda2796$ & G & $-$\tnote{\ddag} & $< 4.05$ & $< 51.89$ \\

$[\mathrm{Ne\,V}]\lambda3426$ & G & $1.22 \pm 0.26$ & $< 3.38$ & $< 5.87$ \\
$[\mathrm{O\,II}]\lambda3726$ & G & $14.09 \pm 0.65$ & $14.57 \pm 1.17$ & $15.91 \pm 2.21$ \\
$[\mathrm{O\,II}]\lambda3729$ & G & $24.83 \pm 0.99$ & $28.41 \pm 1.46$ & $17.14 \pm 2.28$ \\
$[\mathrm{Ne\,III}]\lambda3869$ & G & $23.23 \pm 0.80$ & $42.93 \pm 1.82$ & $43.20 \pm 2.77$ \\
$\mathrm{He\,I}\lambda3889$ & G & $5.29 \pm 0.30$ & $11.92 \pm 0.96$ & $13.58 \pm 1.88$ \\
$\mathrm{H}\epsilon$ & G & $6.70 \pm 0.27$ & $13.29 \pm 0.99$ & $8.75 \pm 1.99$ \\
$\mathrm{H}\delta$ & G & $8.32 \pm 0.21$ & $23.75 \pm 1.15$ & $22.49 \pm 1.94$ \\
$\mathrm{H}\gamma$ & G & $16.29 \pm 0.38$ & $47.05 \pm 1.73$ & $46.11 \pm 2.20$ \\
$[\mathrm{O\,III}]\lambda4363$ & G & $6.21 \pm 0.21$ & $14.16 \pm 1.07$ & $20.21 \pm 1.94$ \\
$\mathrm{He\,II}\lambda4686_\mathrm{narrow}$ & G & $0.76 \pm 0.15$ & $< 4.15$ & $< 5.33$ \\
$\mathrm{He\,II}\lambda4686_\mathrm{broad}$ & G & $< 1.53$ & $< 9.37$ & $< 13.56$ \\
$\mathrm{He\,II}\lambda4686_\mathrm{total}$ & G & $< 2.29$ & $< 13.51$ & $< 18.89$ \\
$\mathrm{H}\beta_\mathrm{narrow}$ & G & $40.32 \pm 3.08$ & $135.38 \pm 17.36$ & $98.01 \pm 7.21$ \\
$\mathrm{H}\beta_\mathrm{broad}$ & G & $13.37 \pm 0.96$ & $< 11.25$ & $36.99 \pm 4.81$ \\
$\mathrm{H}\beta_\mathrm{total}$ & G & $53.60 \pm 3.17$ & $146.63 \pm 0.09$ & $134.42 \pm 6.07$ \\
$[\mathrm{O\,III}]\lambda5007$ & G & $275.01 \pm 26.85$ & $988.73 \pm 203.24$ & $569.72 \pm 44.06$ \\
$\mathrm{He\,I}\lambda5876$ & G & $7.67 \pm 0.23$ & $29.94 \pm 1.00$ & $16.27 \pm 1.90$ \\
$\mathrm{H}\alpha_\mathrm{narrow}$ & G & $185.44 \pm 5.27$ & $814.15 \pm 15.71$ & $309.12 \pm 8.78$ \\
$\mathrm{H}\alpha_\mathrm{broad}$ & G & $212.40 \pm 6.10$ & $223.58 \pm 5.56$ & $538.87 \pm 16.23$ \\
$\mathrm{H}\alpha_\mathrm{total}$ & G & $397.84 \pm 11.31$ & $1037.72 \pm 20.25$ & $847.65 \pm 23.50$ \\
$[\mathrm{N\,II}]\lambda6583$ & G & $22.96 \pm 1.41$ & $< 4.47$ & $< 10.69$ \\
$[\mathrm{S\,II}]\lambda6716$ & G & $8.84 \pm 0.51$ & $10.46 \pm 2.39$ & $< 6.25$ \\
$[\mathrm{S\,II}]\lambda6731$ & G & $10.34 \pm 0.57$ & $8.34 \pm 1.92$ & $< 6.43$ \\
\hline
$\mathrm{Si\,II}\lambda1260$ & P & $-$\tnote{\ddag} & $< 0.51$ & $< 1.61$ \\
$\mathrm{Si\,II}\lambda1304$ & P & $-$\tnote{\ddag} & $< 1.95$ & $< 1.96$ \\
$\mathrm{C\,IV}\lambda1549$\tnote{\dag} & P & $-$\tnote{\ddag} & $< 10.72$ & $< 70.76$ \\
$\mathrm{He\,II}\lambda1640$\tnote{\dag} & P & $-$\tnote{\ddag} & $< 27.47$ & $< 80.58$ \\
$[\mathrm{O\,III}]\lambda1661$ & P & $-$\tnote{\ddag} & $< 4.81$ & $< 8.82$ \\
$[\mathrm{O\,III}]\lambda1666$ & P & $-$\tnote{\ddag} & $< 3.71$ & $< 8.20$ \\
$\mathrm{Si\,III}]\lambda1892$ & P & $4.03 \pm 0.30$ & $4.59 \pm 0.80$ & $< 6.08$ \\
$\mathrm{C\,III}]\lambda1907$ & P & $< 5.22$ & $< 17.59$ & $< 43.08$ \\
$\mathrm{C\,III}]\lambda1909$ & P & $< 4.99$ & $< 16.07$ & $< 39.37$ \\
$[\mathrm{Ne\,IV}]\lambda2424$ & P & $2.88 \pm 0.21$ & $< 1.01$ & $4.18 \pm 1.02$ \\
$\mathrm{O\,II}]\lambda2471$ & P & $1.80 \pm 0.22$ & $< 1.01$ & $6.23 \pm 1.02$ \\
$\mathrm{Mg\,II}\lambda2796$ & P & $7.33 \pm 0.23$ & $3.59 \pm 0.42$ & $< 3.57$ \\
$[\mathrm{Ne\,V}]\lambda3426$ & P & $3.49 \pm 0.28$ & $< 1.82$ & $< 4.39$ \\
$[\mathrm{O\,II}]\lambda3726$ & P & $< 8.71$ & $< 15.22$ & $< 25.84$ \\
$[\mathrm{O\,II}]\lambda3729$ & P & $45.13 \pm 2.93$ & $33.63 \pm 5.11$ & $26.10 \pm 8.59$ \\
$[\mathrm{Ne\,III}]\lambda3869$ & P & $14.11 \pm 0.54$ & $21.42 \pm 1.13$ & $30.22 \pm 2.12$ \\
$\mathrm{He\,I}\lambda3889$ & P & $6.67 \pm 0.41$ & $26.43 \pm 0.97$ & $17.20 \pm 1.57$ \\
$\mathrm{H}\epsilon$ & P & $5.01 \pm 0.21$ & $10.29 \pm 0.62$ & $10.67 \pm 1.40$ \\
$\mathrm{H}\delta$ & P & $2.85 \pm 0.16$ & $13.66 \pm 0.55$ & $22.59 \pm 1.40$ \\
$\mathrm{H}\gamma$ & P & $10.29 \pm 0.35$ & $34.85 \pm 1.10$ & $42.45 \pm 1.81$ \\
$[\mathrm{O\,III}]\lambda4363$ & P & $7.12 \pm 0.33$ & $26.84 \pm 0.99$ & $25.94 \pm 1.60$ \\
$\mathrm{He\,II}\lambda4686$\tnote{\dag} & P & $< 2.60$ & $< 11.19$ & $< 35.40$ \\
$\mathrm{H}\beta$\tnote{\dag} & P & $87.80$ $(\pm 2.15)$ & $158.92 \pm 5.68$ & $114.81 \pm 8.97$ \\
$[\mathrm{O\,III}]\lambda5007$ & P & $242.31 \pm 7.05$ & $1020.11 \pm 12.30$ & $599.57 \pm 13.27$ \\
$\mathrm{He\,I}\lambda5876$ & P & $8.64 \pm 0.23$ & $33.55 \pm 1.00$ & $22.03 \pm 1.99$ \\
$\mathrm{H}\alpha$\tnote{\dag} & P & $263.02 \pm 7.11$ & $< 981.78$ $(\pm 21.61)$ & $703.49 \pm 31.34$ \\
$[\mathrm{N\,II}]\lambda6583$ & P & $< 2.95$ & $< 21.01$ & $< 18.20$ \\
$[\mathrm{S\,II}]\lambda6716$ & P & $< 1.48$ & $23.54 \pm 2.37$ & $< 12.24$ \\
$[\mathrm{S\,II}]\lambda6731$ & P & $7.25 \pm 0.51$ & $< 7.03$ & $< 12.02$ \\
\noalign{\smallskip}
\hline
\end{tabular}
}
\begin{tablenotes}
\item[\dag] {\scriptsize These lines were fitted with narrow+broad components but are reported only as totals, since the prism resolution does not permit a reliable decomposition.}
\item[\ddag] {\scriptsize Lines covered by fewer than 70\% of the individual spectra were excluded from the stacked spectrum and therefore not fitted.}
\end{tablenotes}
\end{threeparttable}
\end{table*}

\section{Accretion Flow Diagnostics at $z>5$ from BLR Emission}

The extreme distances and early cosmic epochs probed by $z > 5$ JWST-selected AGNs offer a unique opportunity to test whether broad-line diagnostics -- developed primarily for low-redshift AGNs or bright QSOs -- can still constrain the physical nature of accretion flows in the low-metallicity, rapidly evolving environments of the early Universe.
To model the emission-line response to different accretion scenarios, we  use of a spectral synthesis code. 
Calculations were performed with version 23.01 of \textsc{Cloudy} \citep{chatzikos2023, Gunasekera2023}. It self-consistently computes the thermal, ionization, and emission-line structure of gas exposed to an AGN continuum.

\subsection{Sub-Eddington Accretion}

\begin{figure}
    \centering
\includegraphics[width=0.9\columnwidth]{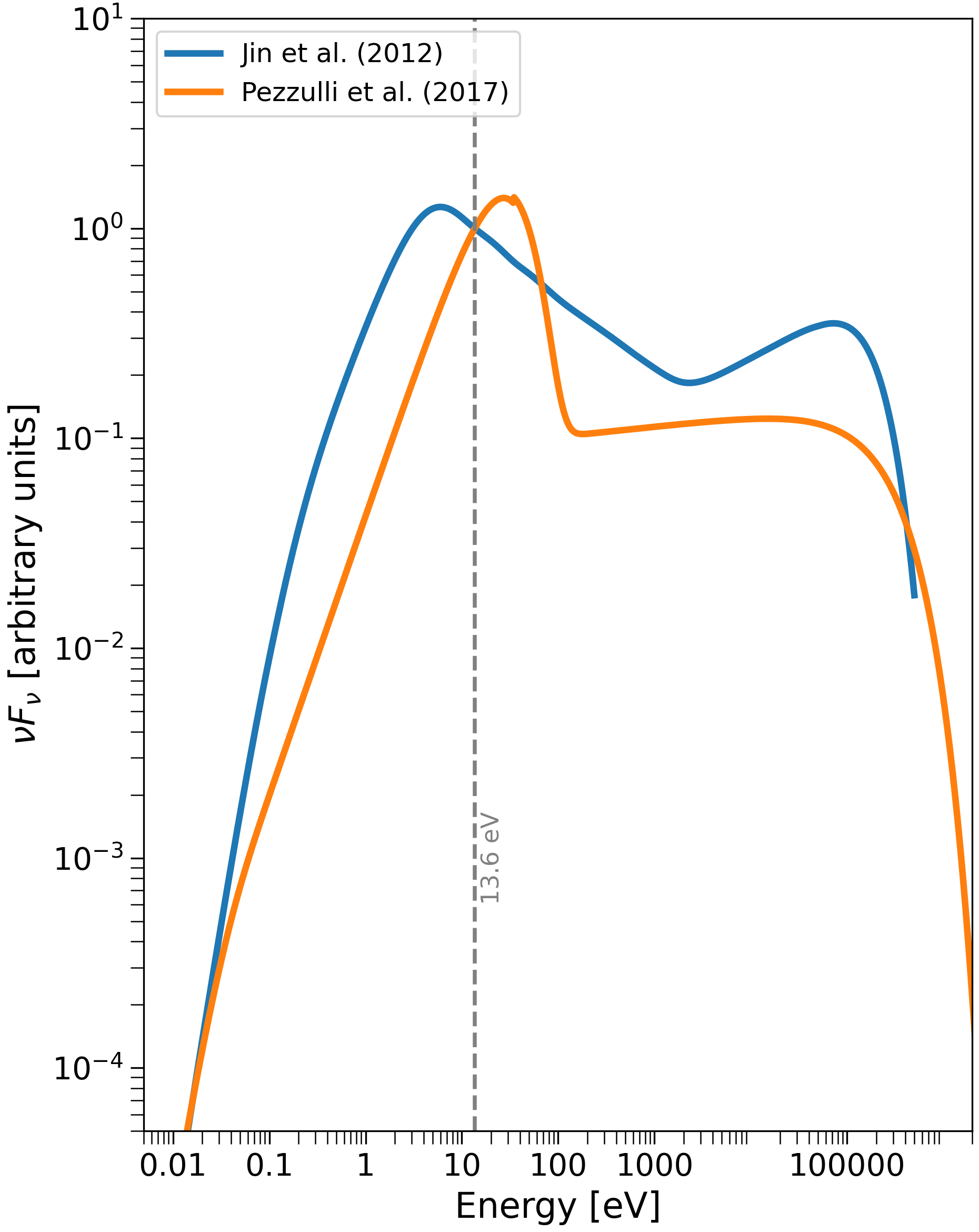}
   \caption{$\nu F_{\nu}$ SEDs of the Jin et al. (2012) and Pezzulli et al. (2017) families, normalized at 13.6 eV. The normalization energy is marked by the dashed vertical line; amplitudes are in arbitrary units and are used solely to compare spectral shapes. The Pezzulli et al. (2017) SED is harder near the hydrogen ionization edge, leading to stronger excitation of high-ionization lines such as He\,II, whereas the Jin et al. (2012) SED exhibits a more prominent soft X-ray tail, contributing extra ionizing photons above the He\,II edge (54.4 eV).}
    \label{fig:sed_jin_pezzulli}
\end{figure}

We focus on sub-Eddington models for the AGN SED, specifically in the context of their ionizing continua. Two representative families of SEDs are considered, whose spectral shapes are compared in Figure~\ref{fig:sed_jin_pezzulli}.

\begin{enumerate}
    \item The empirical SED model of \citet{jin2012a}, constructed from a statistically significant sample of local, unobscured Type 1 AGNs observed with both XMM-Newton and SDSS. This model captures the average broadband emission properties across the optical, UV, and X-ray regimes, and incorporates contributions from the accretion disk, soft excess, and hot corona. Its data-driven nature makes it especially suitable for anchoring the ionizing output of AGNs in the nearby universe under moderate accretion conditions. \\
    
    \item The theoretical framework developed by \citet{pezzulli2017}, which combines a thin accretion disk with a Comptonizing corona to self-consistently model the SED as a function of black hole mass and accretion rate. Designed to probe the growth of early supermassive black holes, this model is tailored for high-redshift AGNs, offering predictions for how the shape and hardness of the ionizing spectrum evolve with redshift and fueling conditions. While the model formally includes photon trapping, this effect only becomes important near the Eddington limit; for sub-Eddington accretion rates, photon trapping is negligible, and the emergent SED reflects the intrinsic disk-corona emission.

\end{enumerate}

\noindent Both SED families were selected to represent sub-Eddington accretion regimes around moderate-mass black holes, with Eddington ratios $L/L_{\rm Edd} \lesssim 0.1$ and typical black hole masses of $10^7$–$10^8\,M_\odot$,
consistent with expectations for radiatively efficient but moderately accreting AGNs \citep[see also the discussion of low $L / L_{\rm edd}$ SED in][]{Ferland2020} and consistent with the average black hole mass of our sample, $\langle M_{\rm BH} \rangle \simeq 10^{7.2} M_\odot$. While single-epoch black hole mass estimates at high redshift might be subject to systematic uncertainties (see e.g. \citet{rusakov_2025,Greene2025}; 
but see e.g. \citet{juodvzbalis2025b} for evidence supporting the locally calibrated single-epoch method), we verified that adopting an AGN SED with a lower BH mass (in our test the \citet{pezzulli2017} SED with \citep[$M_{\rm BH} = 10^6 M_\odot$, i.e., assuming the BH masses were overestimated by 1-2 dex; e.g.,][]{Greene2025} does not change our conclusions. Indeed, as noted by \citet{abuter2024}, single-epoch virial estimators may overestimate black hole masses, but this bias is minimized when using H$\alpha$, for which the error is comparable to the intrinsic scatter of the scaling relation. Also, as recently shown by \citet{juodvzbalis2025b} for an JWST-selected AGN at $z=7$, the BH mass independently measured through kinematics is consistent with the single-epoch mass within 1$\sigma$ scatter.
These templates provide essential input for modeling the ionizing photon budgets and radiative feedback of AGNs during cosmic reionization and structure formation.

\begin{table}[t]
\centering
\begin{threeparttable}
\caption{Physical parameters adopted for the BLR photoionization \textsc{Cloudy} models setup. The two models reported differ only in the ionizing SED.}
\label{tab:cloudy_par}
\begin{tabular}{ll}
\hline
\textbf{Quantity} & \textbf{Adopted value / setting} 
\\
\hline
\noalign{\smallskip}
SEDs\tnote{*} & \cite{jin2012a} \& \cite{pezzulli2017} \\
$n_{\mathrm H}$ & $10^{10}\ \mathrm{cm^{-3}}$ \\
$\log U$ & Grid $-3.5$ to $-1.0$ (step 0.5 dex) \\
Covering factor & 1 \\
Pressure & Constant (isobaric) \\
$Z$ & $0.1\,Z_\odot$ \\
N/O prescription & \cite{groves2004} \\
He prescription & \citet{dopita2000} \\
Solar ref. & \citet{grevesse2010} \\
Dust & None (grains disabled) \\
Molecules & Disabled \\
$v_{\mathrm{turb}}$ & $100\ \mathrm{km\,s^{-1}}$ \\
$N_{\mathrm H}$ stop & $10^{23}\ \mathrm{cm^{-2}}$ \\
$T_{\mathrm{stop}}$ & $4000\ \mathrm{K}$ \\
$x_e$ stop & $10^{-2}$ \\
H-like levels & 10 resolved \\
Convergence & Iterated to convergence \\
\hline
\end{tabular}
\begin{tablenotes}
\item[*] \citet{jin2012a}: $M_{\rm BH}=10^{7}-10^{8}M_\odot$, $\lambda_{\rm edd} = 0.07$;\\ \citet{pezzulli2017}: $M_{\rm BH}=10^{7} M_\odot$, $\lambda_{\rm edd} = 0.1$.
\end{tablenotes}
\end{threeparttable}
\end{table}

We constructed a grid of photoionization models tailored to conditions typical of the BLR (parameters are summarized in Table~\ref{tab:cloudy_par}). All models assume constant-pressure conditions and a metallicity of $0.1\,Z_\odot$, broadly consistent with what inferred for these AGNs at $z \gtrsim 5$ \citep[see][]{maiolino2024b,trefoloni_2025}. We adopted the solar reference from \citet{grevesse2010} with variable He/H, C/O, and N/O according to the metallcity. Helium enrichment follows a primary-production from \citet{dopita2000}, and nitrogen and carbon follow a primary+secondary prescription from \citet{groves2004}. Dust and molecules were excluded, consistent with the BLR being within the sublimation radius. Each cloud was truncated at $\log(N_{\mathrm{H}}/{\rm cm}^{-2}) = 23$, and models were terminated when the electron fraction fell below $10^{-2}$ or the temperature dropped below 4000 K. The hydrogen gas density was fixed at $\log\,(n_{\rm H}/{\rm cm^{-3}}) = 10$, and the ionization parameter, i.e. the dimensionless ratio between the density of hydrogen-ionizing photons and the hydrogen number density, was varied across the range $\log U = -3.5$ to $-1.0$. The inclusion of low ionization parameters (down to $\log U \simeq -3.5$) allowed us to probe BLR conditions where the number of hydrogen-ionizing photons per particle is minimal.
At fixed gas density, decreasing $\log U$ causes the zone of doubly ionized helium (HeIII) and other high‑ionization regions to contract, thereby reducing the luminosity of recombination lines such as HeII$\lambda$4686 and high-ionization collisionally excited lines such as CIV$\lambda$1549. In contrast, low-ionization lines (e.g., H$\beta$) remain less affected or may even increase in relative strength. If the observed weakness of high-ionization features is primarily driven by a low ionization parameter, models with $\log U \lesssim -3$ should reproduce the observed H$\beta$/HeII$\lambda4686$ broad-line ratios. Failure to reproduce these ratios at low $\log U$ would instead support alternative explanations.

In parallel, we compared the EWs of H$\beta$, HeII\,$\lambda4686$, and CIV\,$\lambda1549$ predicted by the models to those observed. In \textsc{Cloudy}, EWs were computed by integrating the emergent line flux and referencing it to the transmitted continuum at the line wavelength. To isolate geometric effects, we ran models across a range of BLR covering factors (CFs). If the observed EWs of all broad emission lines are reproduced by a common CF envelope, the variations can be attributed to a non-unity covering factor rather than intrinsic differences in the ionizing continuum.
Conversely, if the observed values require different CFs for different lines, or lie outside the model envelope altogether, then a purely geometric interpretation is insufficient. This would instead point to intrinsic shortcomings in the assumed SEDs, ionization structure, or BLR physics. For upper and lower limits, we consider models to be consistent if their CF envelope intersects the region allowed by the observations. A persistent mismatch across the full ionization grid would favor scenarios involving a soft or filtered SED, or super-Eddington accretion.

\begin{figure}[!htb]
    \centering
    \includegraphics[width=\columnwidth]{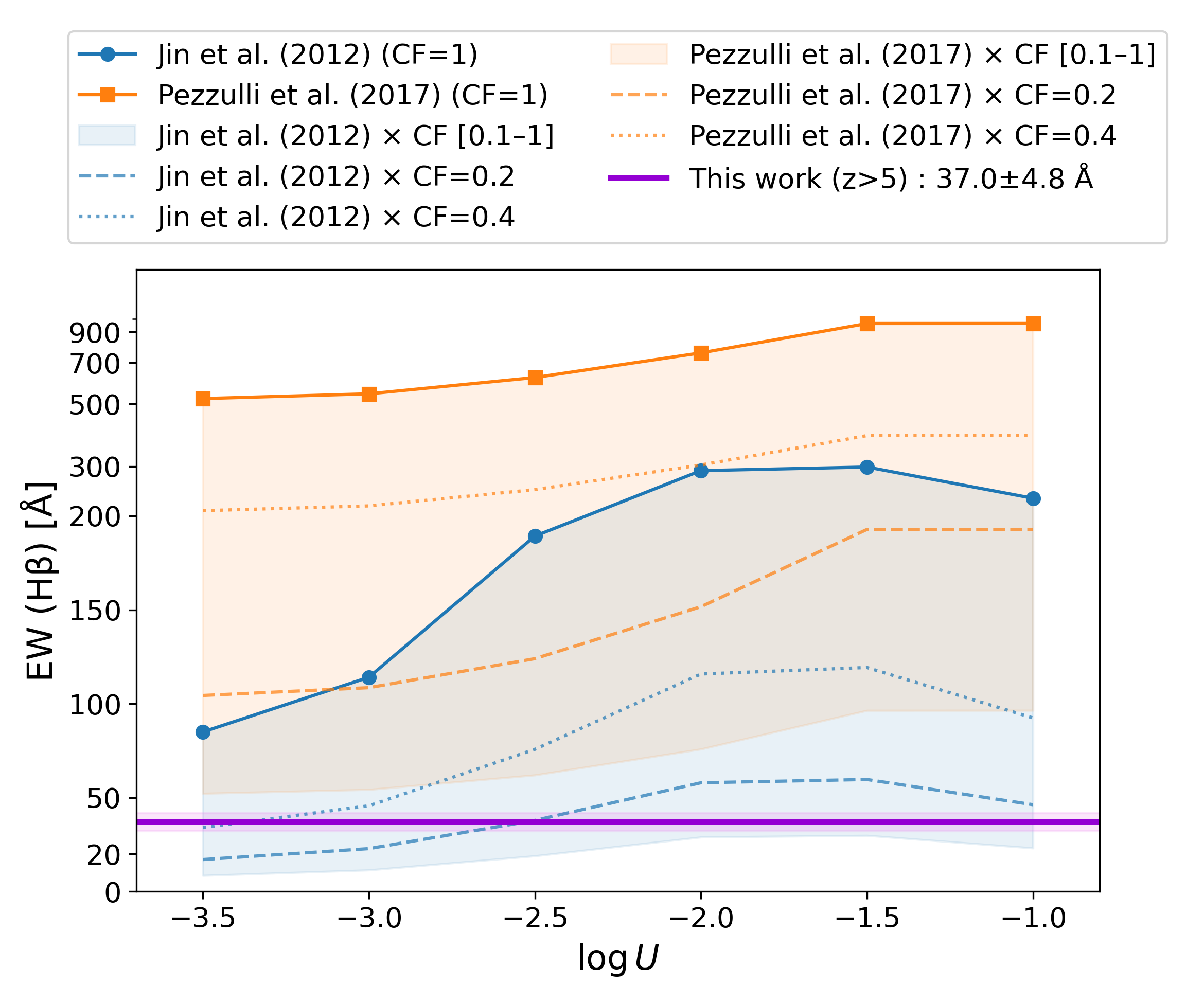}
    \caption{Broad H$\beta$ EWs as a function of ionization parameter $\log U$, derived from sub-Eddington \textsc{Cloudy} models (parameters listed in Table~\ref{tab:cloudy_par}). Different line styles indicate varying broad-line region CFs, ranging from 1.0 to 0.1; models with the same SED share the same color. The shaded region for each SED represents the envelope spanned by CF values from 1.0 down to 0.1. The dark violet horizontal line marks the observed EW of broad H$\beta$ measured in the $z > 5$ grating stack. The vertical axis uses a symlog scale (linear below 200 \AA, logarithmic above).
    }
\label{fig:cloudy_subedd_ew_hbeta}
\end{figure}

\begin{figure}[!htb]
    \centering
    \includegraphics[width=\columnwidth]{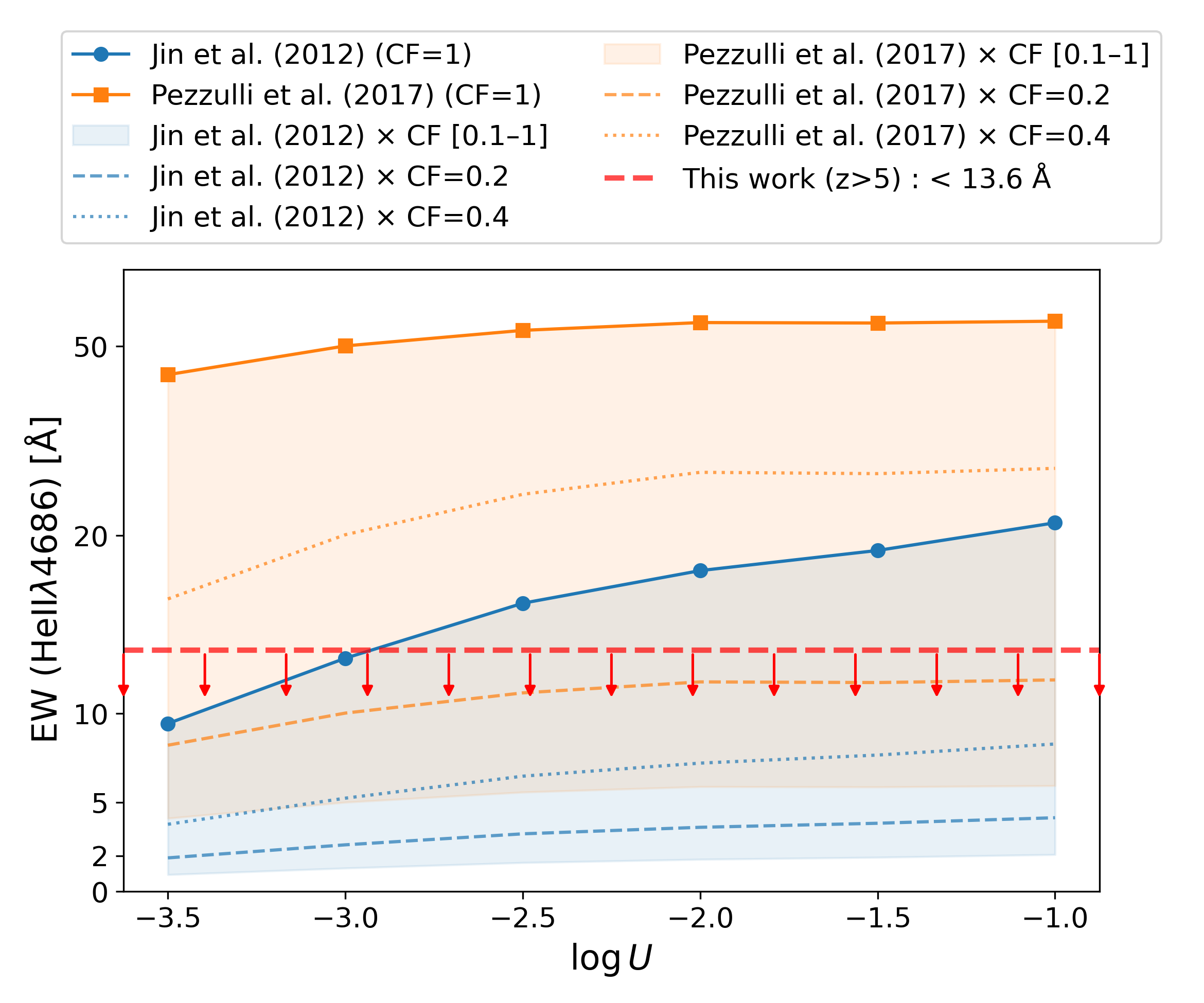}
   \caption{Broad HeII$\lambda4686$ EWs as a function of ionization parameter $\log U$, derived from sub-Eddington \textsc{Cloudy} models (parameters listed in Table~\ref{tab:cloudy_par}). Different line styles indicate varying broad-line region CFs, from 1.0 to 0.1; models with the same SED share the same color. The shaded region for each SED shows the envelope spanned by varying CF. Red downward arrows indicate the $3\sigma$ upper limit on the EW of broad HeII$\lambda4686$ measured in the $z > 5$ grating stack. The vertical axis uses a symlog scale (linear below 25 \AA, logarithmic above). }
    \label{fig:cloudy_subedd_ew_heii}
\end{figure}

\begin{figure}[!htb]
    \centering
    \includegraphics[width=\columnwidth]{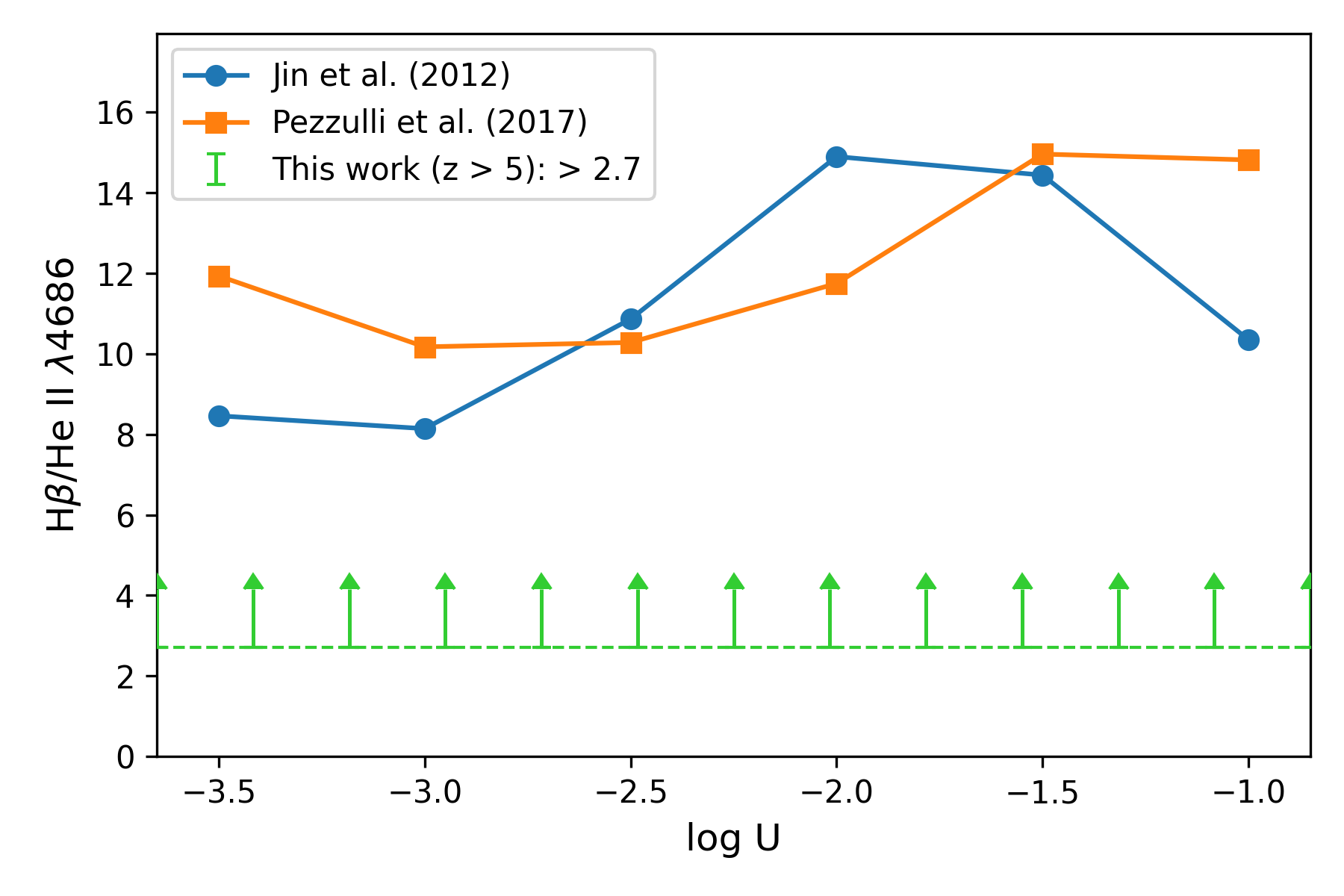}
    \caption{Predicted ratio of broad H$\beta$ to HeII$\lambda4686$ emission as a function of ionization parameter $\log U$, based on the sub-Eddington \textsc{Cloudy} models listed in Table~\ref{tab:cloudy_par}. Blue and orange curves correspond to the Jin et al. and Pezzulli et al. SEDs, respectively. As the ratio is independent of covering factor, only one curve is shown per SED. The green upward arrow marks the $3\sigma$ lower limit on the broad-line H$\beta$/HeII ratio measured in the $z > 5$ grating stack.}
    \label{fig:cloudy_hbeta_heii}
\end{figure}

\begin{figure}[!htb]
    \centering
    \includegraphics[width=\columnwidth]{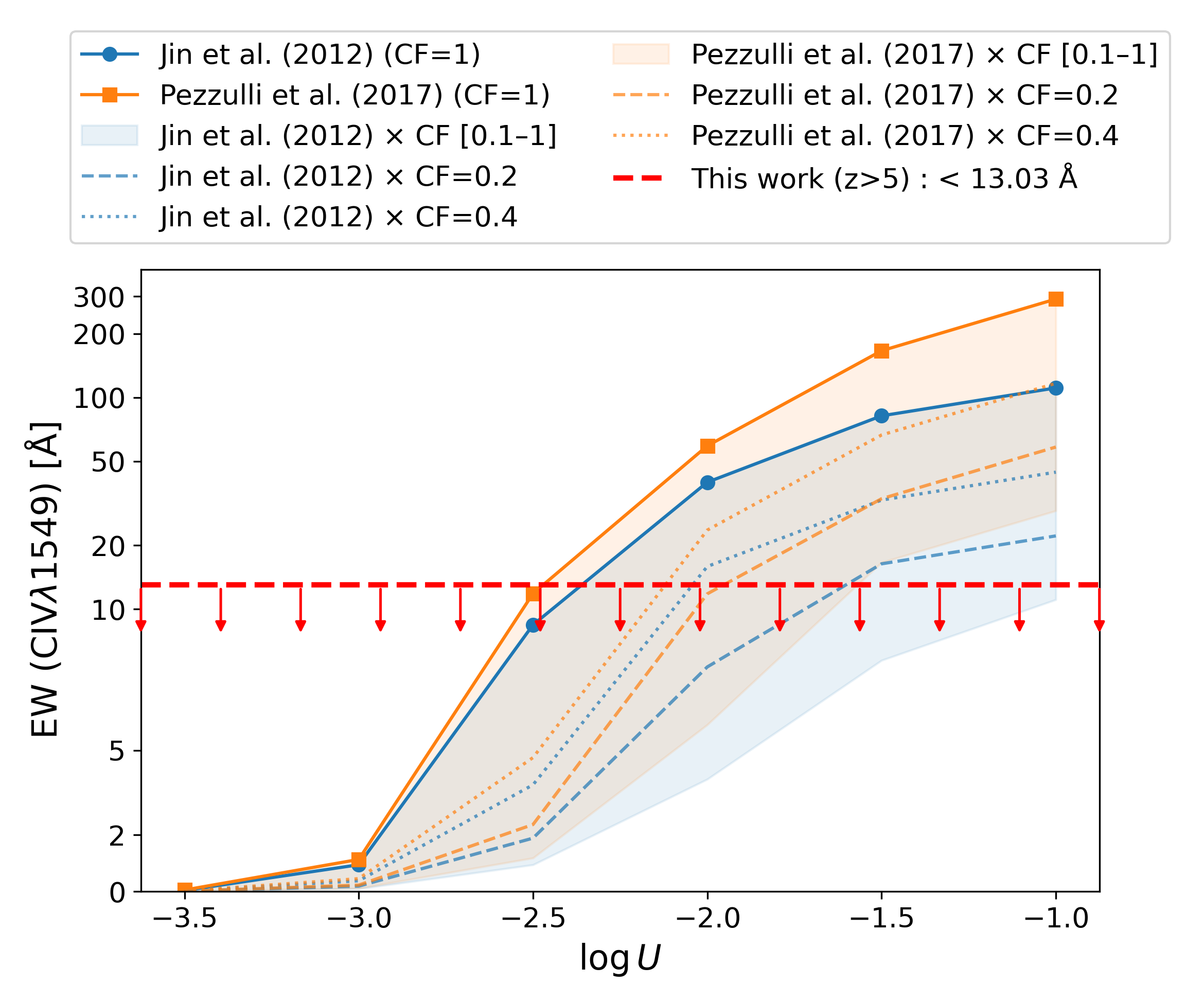}
   \caption{Broad C IV $\lambda1549$ EWs vs. ionization parameter $\log U$ from sub-Eddington \textsc{Cloudy} models (parameters in Table~\ref{tab:cloudy_par}). Line styles indicate different broad-line region CFs; colors distinguish SEDs. Red downward arrows mark the $3\sigma$ upper limit from the $z>5$ grating stack. For consistency with our measurements, only the stronger CIV$\lambda1549$ component from \textsc{Cloudy} is shown. The vertical axis uses a symlog scale (linear below 10 \AA, logarithmic above).}
    \label{fig:cloudy_subedd_ew_civ}
\end{figure}

Figures~\ref{fig:cloudy_subedd_ew_hbeta}–\ref{fig:cloudy_subedd_ew_civ} compare the model predictions from the two sub-Eddington SED templates against four key spectroscopic diagnostics in the $z>5$ stack:

\medskip
\begin{enumerate}

\item \textbf{H$\beta$ Equivalent Width} (Fig. \ref{fig:cloudy_subedd_ew_hbeta}).
The observed H$\beta$ EW of $\sim$37\,\AA\ is matched only by the Jin et al. soft SED, and only within a narrow region at low ionization ($\log U \lesssim -3$) and low covering factor (CF $\lesssim 0.2$). For higher $\log U$, the predicted EW exceeds the observed value unless the CF is reduced to unrealistically low levels. The Pezzulli et al. SED fails to reproduce the observed H$\beta$ strength at any $\log U$ or CF combination, consistently overpredicting the line flux even at minimal covering. These constraints strongly disfavor harder SEDs and point to the need for both soft ionizing spectra and small BLR covering factors.

\medskip
\item \textbf{HeII$\lambda$4686 Equivalent Width} (Fig. \ref{fig:cloudy_subedd_ew_heii}).
The $3\sigma$ upper limit on the broad HeII EW ($\sim$13\,\AA) rules out the Pezzulli et al. SED for models with CFs $\gtrsim 0.2$, where it predicts HeII strengths exceeding 15\,\AA. The Jin et al. model, with a softer ionizing continuum, yields substantially lower HeII EWs and remains consistent with the observational upper limit across the entire $\log U$–CF space (except for the highest CFs at $\log U \gtrsim -3$). These results disfavor hard, unfiltered SEDs unless the BLR is shielded from high-energy photons, such as in geometries with inner absorption or anisotropic emission.

\medskip
\item \textbf{H$\beta$/HeII$\lambda$4686 Ratio} (Fig. \ref{fig:cloudy_hbeta_heii}).
The observed lower limit on the broad H$\beta$/HeII ratio ($> 2.7$) is met by both models across the entire $\log U$ space. While this diagnostic does not strongly distinguish between the two SEDs, it supports the same region in $\log U$–CF space favored by the individual line EWs.

\medskip
\item \textbf{CIV$\lambda$1549 Equivalent Width} (Fig. \ref{fig:cloudy_subedd_ew_civ}).
The $3\sigma$ upper limit of $\sim$13\,\AA\ on the CIV EW provides stringent constraints. Both the Pezzulli et al. and Jin et al. SED overpredicts CIV emission at $\log U \gtrsim -2.5\,$–$\,{-2}$ and CF $\gtrsim 0.4$, and are therefore strongly disfavored in that regime.
\end{enumerate}

\noindent Together, these four diagnostics favor a model with a soft ionizing continuum (e.g., Jin et al.), moderate ionization parameter ($\log U \sim -2.5$), and a covering factor of $\sim 0.1$-$0.5$. Harder SEDs such as Pezzulli et al. are incompatible with the combined observational constraints unless implausibly low BLR covering factors are invoked. 

\section{Summary and Discussion}

We have presented a spectroscopic analysis of 32 broad-line active galactic nuclei (AGNs) identified in JWST/NIRSpec observations from the JADES survey, spanning a redshift range $1.7 < z < 9$. The initial parent sample contained 34 objects, but two X-ray–detected AGNs were excluded from the present analysis. Emission-line diagnostics from both prism and grating stacks were used to characterize the ionizing spectra and physical conditions of the BLR, with particular focus on the high-redshift ($z > 5$) regime, where AGNs show puzzling deviations from standard photoionization expectations. While equivalent widths for all redshift bins are listed in Table \ref{tab:ew}, Figures 5–8 focus exclusively on the $z>5$ stacks and their corresponding photoionization models. This work is primarily concerned with the high-redshift AGN population; the lower-redshift stacks ($z<3.5$, based on only five objects, and $3.5<z<5$) are included mainly for completeness rather than detailed analysis. Moreover, these data are not constraining, and their inclusion in the plots would not yield any additional insight. A comparison between models and these lower-redshift bins is beyond the scope of this study. Consequently, our discussion and conclusions are restricted to the $z>5$ sample, and equivalent widths from the lower-redshift stacks are not used in any physical interpretation. 

At $z > 5$, we find a suppression of high-ionization rest-frame UV lines such as HeII$\lambda$1640, CIV$\lambda$1549, and NV$\lambda$1240. These lines are either undetected or appear with extremely low EWs, while Balmer lines (H$\alpha$, H$\beta$) and low-ionization lines like HeI$\lambda$5876 remain strong and well detected. Narrow forbidden lines like [NeV]$\lambda$3426 (ionization potential $\sim 97$ eV) and [NeIV]$\lambda$2424 ($\sim 63$ eV) are also weak or absent at high redshift, further supporting a soft ionizing photon field.

To interpret these trends, we constructed a suite of \textsc{Cloudy} photoionization models using two sub-Eddington spectral energy distributions (SEDs): the empirical template of \citet{jin2012a}, representative of nearby AGNs with moderate Eddington ratios, and the theoretical disk-corona model of \citet{pezzulli2017}, designed for high-redshift black hole growth. We varied the ionization parameter $\log U$ from $-3.5$ to $-1.0$ and explored a range of CFs between 0.1 and 1.0. All models assumed constant-pressure clouds with $n_{\mathrm{H}} = 10^{10}\,\mathrm{cm^{-3}}$ and $Z = 0.1\,Z_\odot$, and included realistic prescriptions for helium, nitrogen, and carbon enrichment. Dust and molecules were excluded.

Four key diagnostics were compared against the $z > 5$ observations: the EWs of H$\beta$, HeII$\lambda$4686, and CIV$\lambda$1549, and the H$\beta$/HeII$\lambda$4686 broad-line ratio. These constraints jointly delineate a narrow region of viable model space: 1) The observed H$\beta$ EW ($\sim 37$\,\AA) is never matched by the Pezzulli et al. SED across the full $\log U$ and CF space. The Jin et al. SED reproduces the observed value only at very low ionization ($\log U \sim -3$) and low covering factors (CF $\sim 0.1$--$0.3$). This indicates a strong preference for soft ionizing continua and limited BLR coverage; 2) The $3\sigma$ upper limit on HeII ($\sim 13$\,\AA) excludes the Pezzulli et al. model at $\log U \gtrsim -2.5$, where predicted EWs exceed 15\,\AA\ even at CF $\gtrsim 0.2$. The softer Jin et al. SED predicts significantly lower HeII emission and remains fully consistent with the observational limit across the full $\log U$-CF parameter space. These results disfavor hard SEDs unless the BLR is substantially shielded from high-energy photons; 3) The observed lower limit on the broad H$\beta$/HeII ratio ($> 2.7$) is not a strong discriminator between SEDs and is satisfied by both models across the whole $\log U$ space; and 4) The 3$\sigma$ upper limit of $\sim$13\,\AA\ on the C\,IV equivalent width provides a stringent constraint. Both the Pezzulli et al. and Jin et al. SEDs tend to overpredict C\,IV emission for $\log U \gtrsim -2$ and $\mathrm{CF} \gtrsim 0.4$, and are therefore disfavored in that region of parameter space.

Taken together, these four diagnostics favor a model with soft ionizing continuum (e.g., Jin et al.), moderate-to-low ionization parameter ($\log U \sim -2.5$ to $-3$), and a covering factor of $\sim 0.1$--$0.5$. Harder SEDs such as Pezzulli et al.\ are incompatible with the combined observational constraints unless implausibly low BLR covering factors are invoked. 
We verified that these conclusions are not sensitive to the adopted gas density. We performed additional tests using photoionization models with $\log\,  (n_{\rm H}/{\rm cm}^{-3}) = 8$ and 12. While varying $n_{\rm H}$ shifts the absolute equivalent width values and can slightly modify the preferred ionization parameter $\log U$, it does not resolve the key discrepancies between model predictions and observed data.
Even within the preferred regime, some tension remains in simultaneously reproducing all observed EWs, highlighting possible limitations of static, sub-Eddington photoionization models.
These discrepancies may motivate the need for alternative physical scenarios. One such possibility is the presence of a filtered or anisotropic ionizing continuum, as may occur in super-Eddington accretion flows. In this case, the far-UV and soft X-ray continuum can be shielded from the BLR by inner disk structures or collimated outflows, selectively suppressing high-ionization lines while leaving Balmer emission relatively unaffected. This interpretation is consistent with both the observed line EWs and the known X-ray weakness of many high-$z$ JWST AGNs.

Super-Eddington accretion flows are expected to yield emergent spectra that differ markedly from the standard thin-disk paradigm. Recent calculations by \citet{Madau2025} model the radiative output of such flows, incorporating the effects of radiation pressure and funnel-like geometries. In this framework, the inner disk inflates to form an optically thick funnel along the rotation axis, collimating ionizing photons toward polar directions. Equatorial sightlines are dominated by reprocessed emission, resulting in highly anisotropic SEDs: the escaping radiation becomes significantly harder and more luminous at low inclination angles, while the extreme UV and soft X-ray output is strongly suppressed for $\theta \gtrsim 60^\circ$.
This anisotropy has important consequences for emission-line diagnostics. The angular dependence of the ionizing continuum leads to a reduced production of HeII-ionizing photons at large viewing angles, even though the hydrogen-ionizing luminosity remains high when averaged over solid angle. Such directionally filtered SEDs may provide a compelling explanation for the observed combination of strong Balmer lines and weak high-ionization emission in many JWST-selected broad-line AGNs. Detailed implications for BLR line ratios and covering factors will be explored in a companion paper.

An alternative explanation for the observed suppression of high-ionization lines in JWST AGNs invokes obscuration or filtering of the ionizing continuum by intervening material. \citet{Tang2025} suggest that the absence of lines such as HeII and CIV in many high-redshift sources may from attenuation by dense neutral gas and dust, which absorb the EUV and soft X-ray photons required to excite these transitions. Similarly, \citet{Ishibashi2025} propose that X-ray weakness arises from radiative feedback and obscuration by dust-poor, high-column-density gas, which can absorb high-energy photons while allowing lower-energy emission (e.g., Balmer lines) to escape. These scenarios imply that the BLR may not be illuminated uniformly, and that anisotropic or filtered radiation fields -- rather than intrinsically soft SEDs or extreme accretion physics -- could account for the observed spectroscopic trends. Future constraints on the geometry, composition, and ionization state of the absorbing medium will be key to distinguishing between these possibilities.

Our photoionization analysis assumes that the observed EWs directly reflect the underlying ionizing SED, with no differential dust extinction between the BLR and the continuum source. If the dust acts as a uniform screen at large distances, EWs remain unchanged and our {\sc Cloudy} modeling remains valid. However, the observed Balmer decrements (H$\alpha$/H$\beta > 10$) suggest significant reddening (see Table \ref{tab:ew} and \citealt{Brooks2025}), which could imply non-negligible wavelength-dependent attenuation. If dust is patchy or mixed within the BLR, or if it affects line and continuum emission differently, then our assumptions may break down.
Alternatively, the large broad-line Balmer decrement could be a result of strong collisional excitation in the BLR, as observed and suggested for some lower redshift AGNs \citep{ilic_2012}, or radiative transfer effect when Balmer lines become optically thick \citep{Chang_2025,ji_locallrd_2025}.
We will explore alternative BLR models to investigate the above scenarios in future work.

\begin{acknowledgement}
    XJ and RM acknowledge ERC Advanced Grant 695671 “QUENCH” and support by the Science and Technology Facilities Council (STFC) and by the UKRI Frontier Research grant RISEandFALL.
    This work is based on observations made with the NASA/ESA/CSA James Webb Space Telescope. The data are available at the Mikulski Archive for Space Telescopes (MAST) at the Space Telescope Science Institute, which is operated by the Association of Universities for Research in Astronomy, Inc., under NASA contract NAS 5-03127 for JWST.
    The reduced spectra used in this work are from the public data release of the JADES survey, available at https://jades-survey.github.io/scientists/data.html.
    
\end{acknowledgement}

%-------------------------------------------------------------------
\bibliographystyle{aa}
\bibliography{biblio}

%---------------------------

\clearpage
\appendix
\onecolumn
\section{Additional Material}
\label{appendix}
\vspace{1cm}

\begin{figure}[htb]
    \centering
    \includegraphics[width=0.99\textwidth]{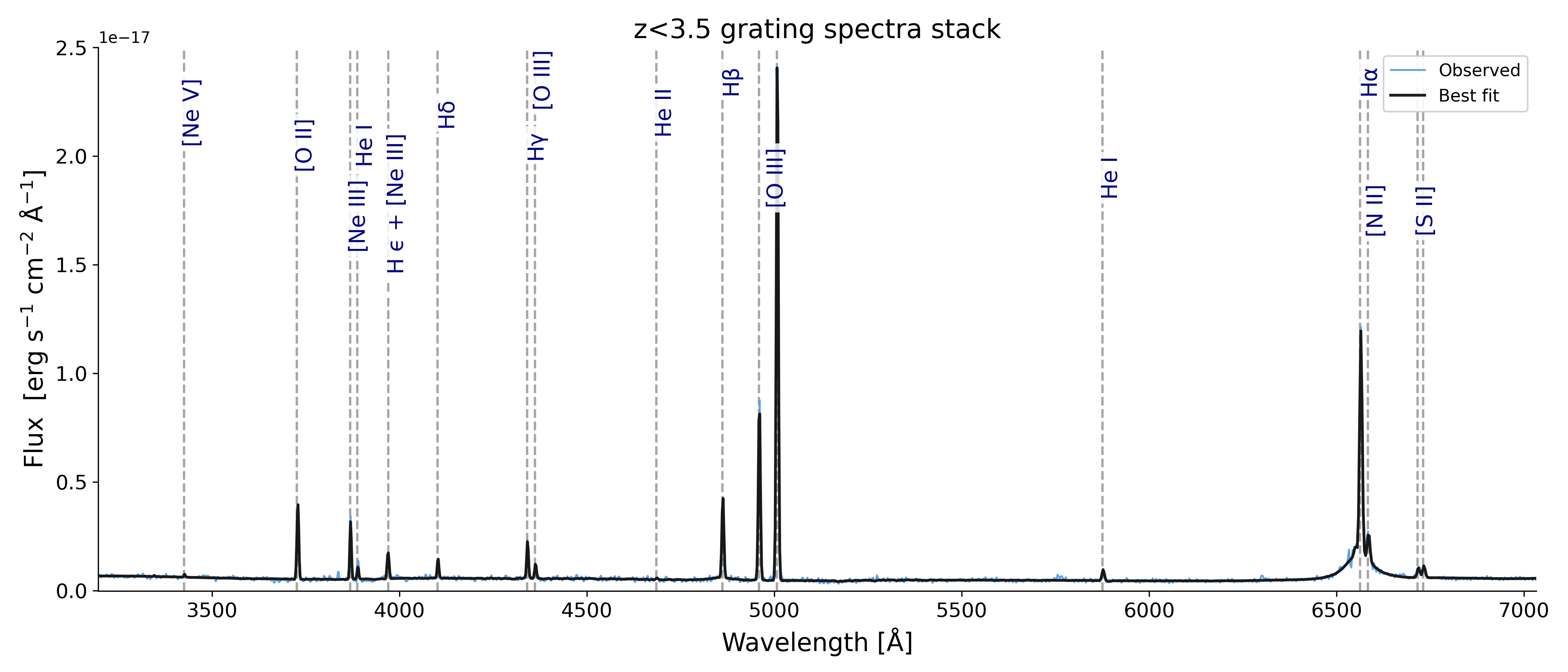}
    \caption{Stack of the rest-frame (G140M+G235M+G395M) grating spectra in the low-redshift bin ($z < 3.5$). The black line represents the best-fit from our \texttt{pPXF} fitting procedure, while the blue line shows the observed stacked spectrum. For visualization, the normalized stack was multiplied by the mean [OIII]$\lambda$5007 flux of the 5 contributing objects; this rescaling is used only for the figure and does not affect any measurements.}
    \label{fig:z35_grating_stack}
\end{figure}

\vspace{2cm}
\begin{figure}[htb]
    \centering
    \includegraphics[width=0.99\textwidth]{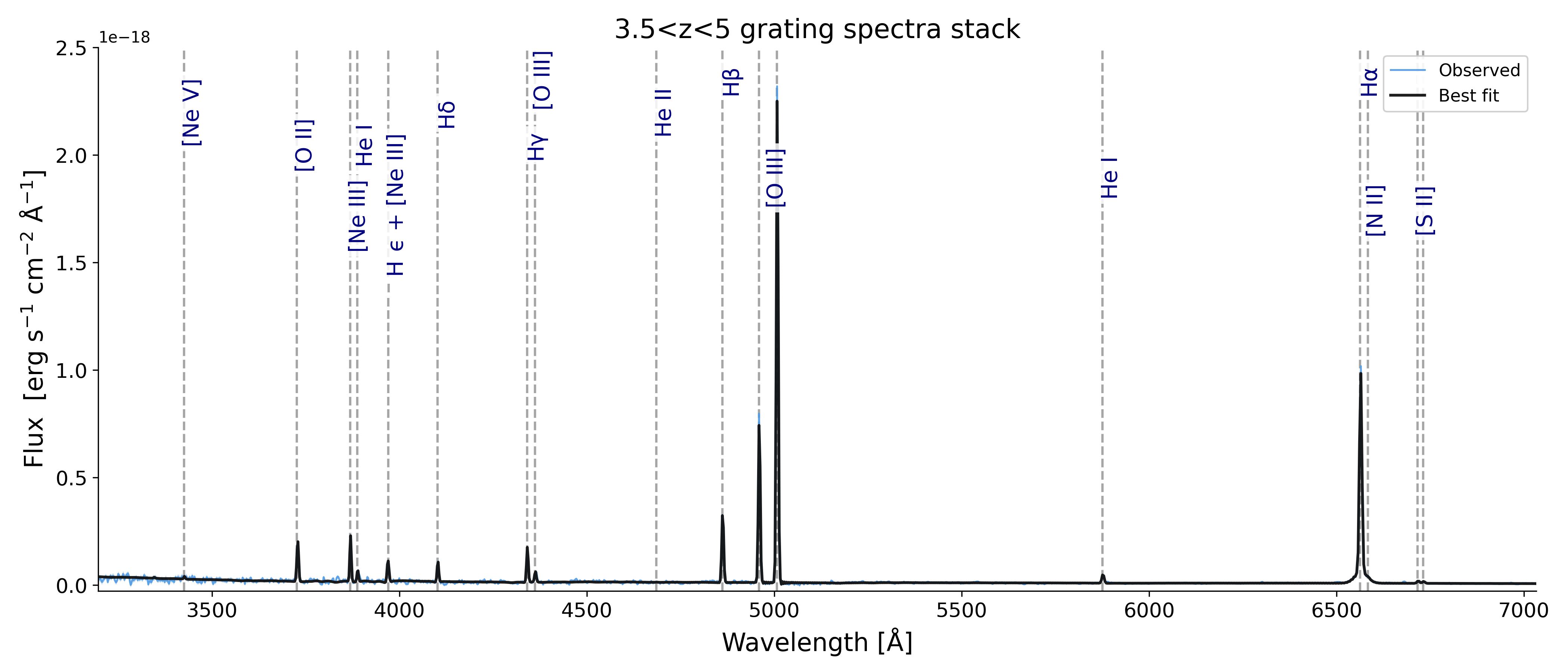}
    \caption{Stack of the rest-frame (G140M+G235M+G395M) grating spectra in the intermediate redshift bin ($3.5 < z < 5$). Color coding as figure above. For visualization purposes, the normalized stack has been multiplied by the mean [OIII]$\lambda$5007 flux of the 12 contributing objects. This rescaling affects only the plotted figure and not the underlying measurements.}
    \label{fig:z35_z5_grating_stack}
\end{figure}

\begin{figure}[htb]
    \centering
    \vspace{1cm}
    \includegraphics[width=0.99\textwidth]{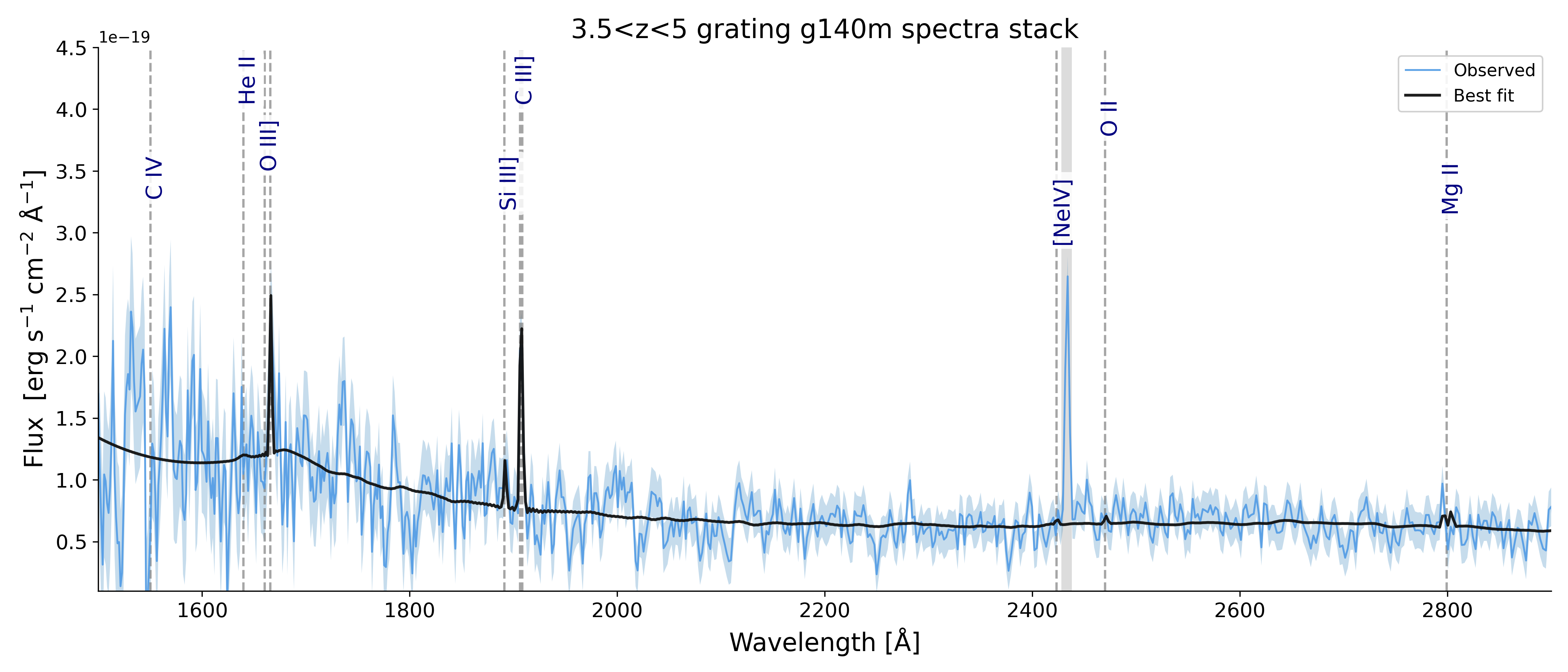}
    \caption{Stack of the rest-frame G140M grating spectra in the intermediate-redshift bin ($3.5 < z < 5$). Color coding as figures above. For visualization, the normalized stack was multiplied by the mean [OIII]$\lambda$5007 flux of the 11 contributing objects; this rescaling is used only for the figure and does not affect any measurements. The gray band masks a probable oversampling error, thus it has been removed from the fitting.}
    \label{fig:z35_z5_g140m_stack}
\end{figure}

\vspace{2cm}
\begin{figure}[htb]
    \centering
    \includegraphics[width=0.99\textwidth]{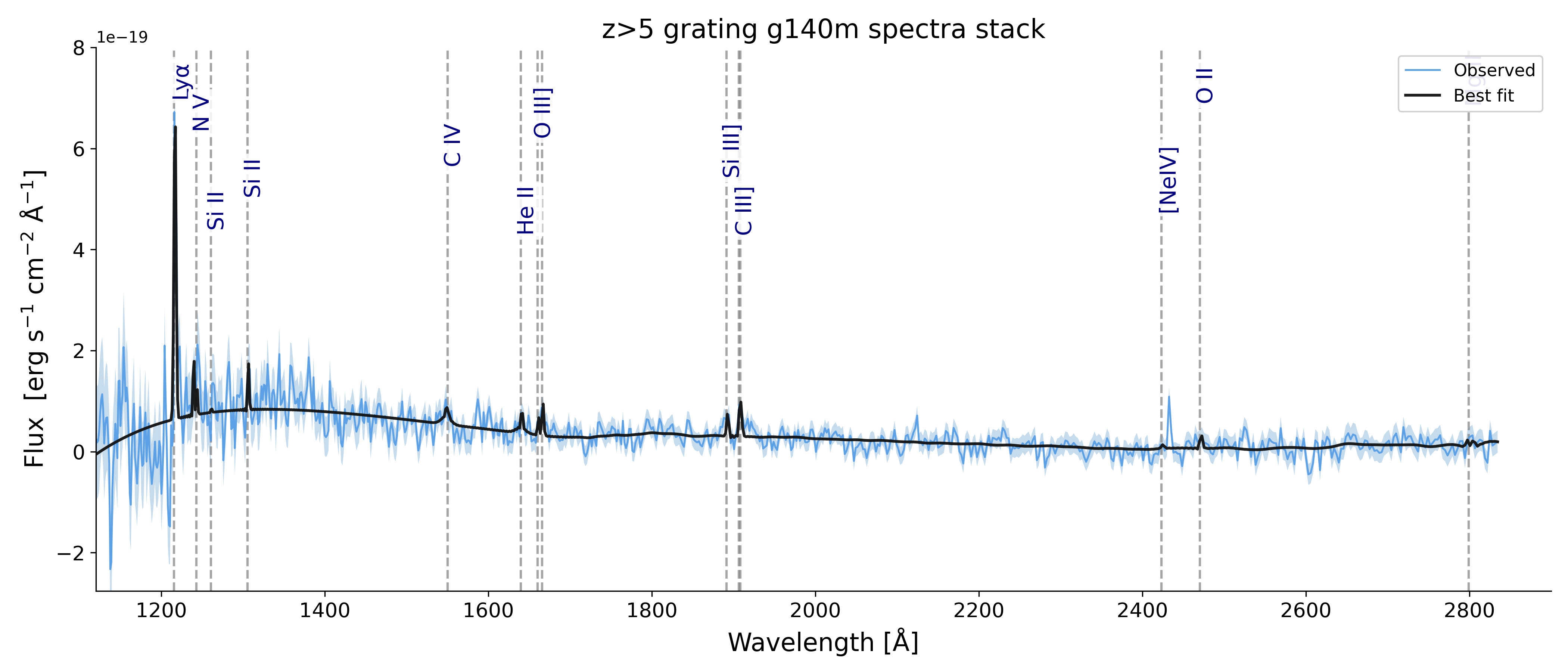}
    \caption{Stack of the rest-frame G140M grating spectra in the high-redshift bin ($z > 5$). Color coding as figures above. For visualization, the normalized stack was multiplied by the mean [OIII]$\lambda$5007 flux of the 14 contributing objects; this rescaling is used only for the figure and does not affect any measurements.}
    \label{fig:z5_g140m_stack}
\end{figure}

\begin{figure}[htb]
    \centering
\vspace{1cm}
\includegraphics[width=0.99\textwidth]{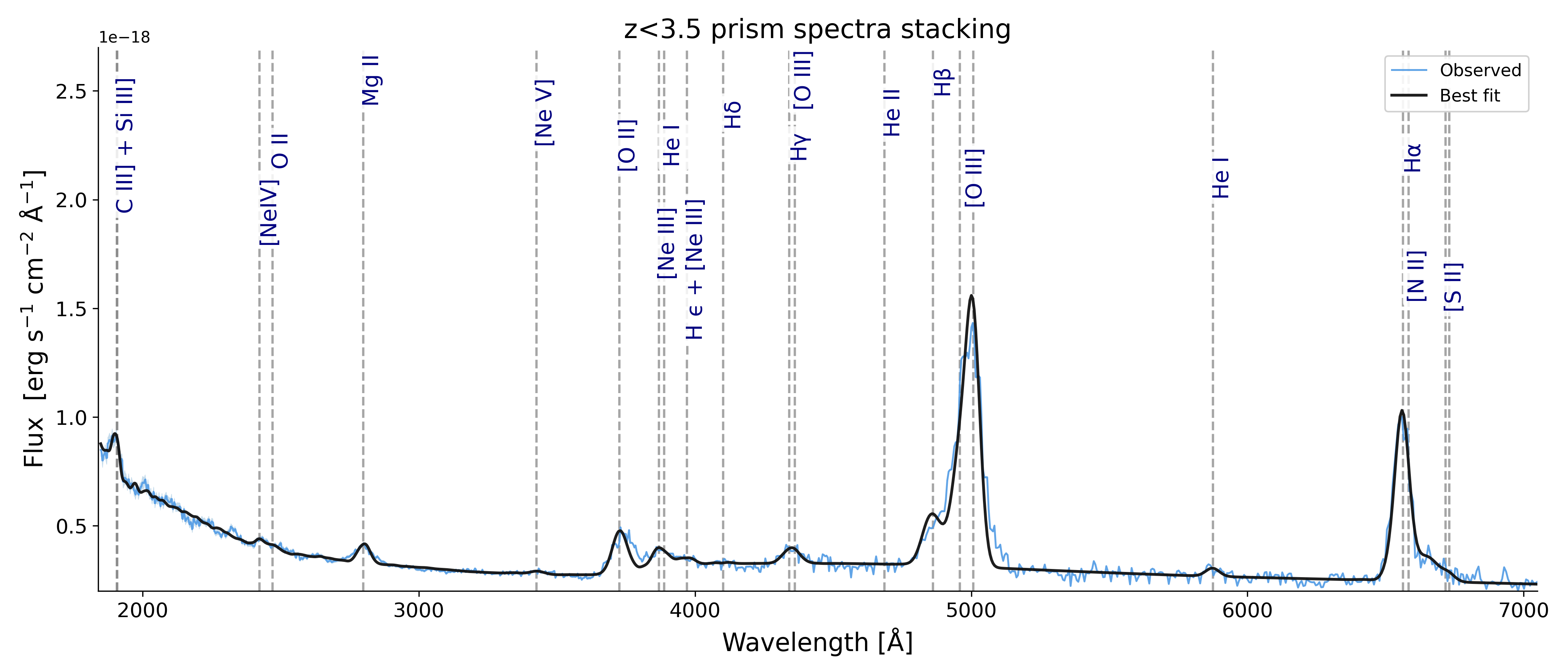}
    \caption{Stack of the rest-frame prism spectra in the low-redshift bin ($ z < 3.5$). Color coding as figures above. For visualization purposes, the normalized stack has been multiplied by the mean [OIII]$\lambda$5007 flux of the 5 contributing objects. This rescaling affects only the plotted figure and not the underlying measurements.}
    \label{fig:z35_prism_stack}
\end{figure}

\vspace{2cm}
\begin{figure}[htb]
    \centering
\vspace{1cm}
\includegraphics[width=0.99\textwidth]{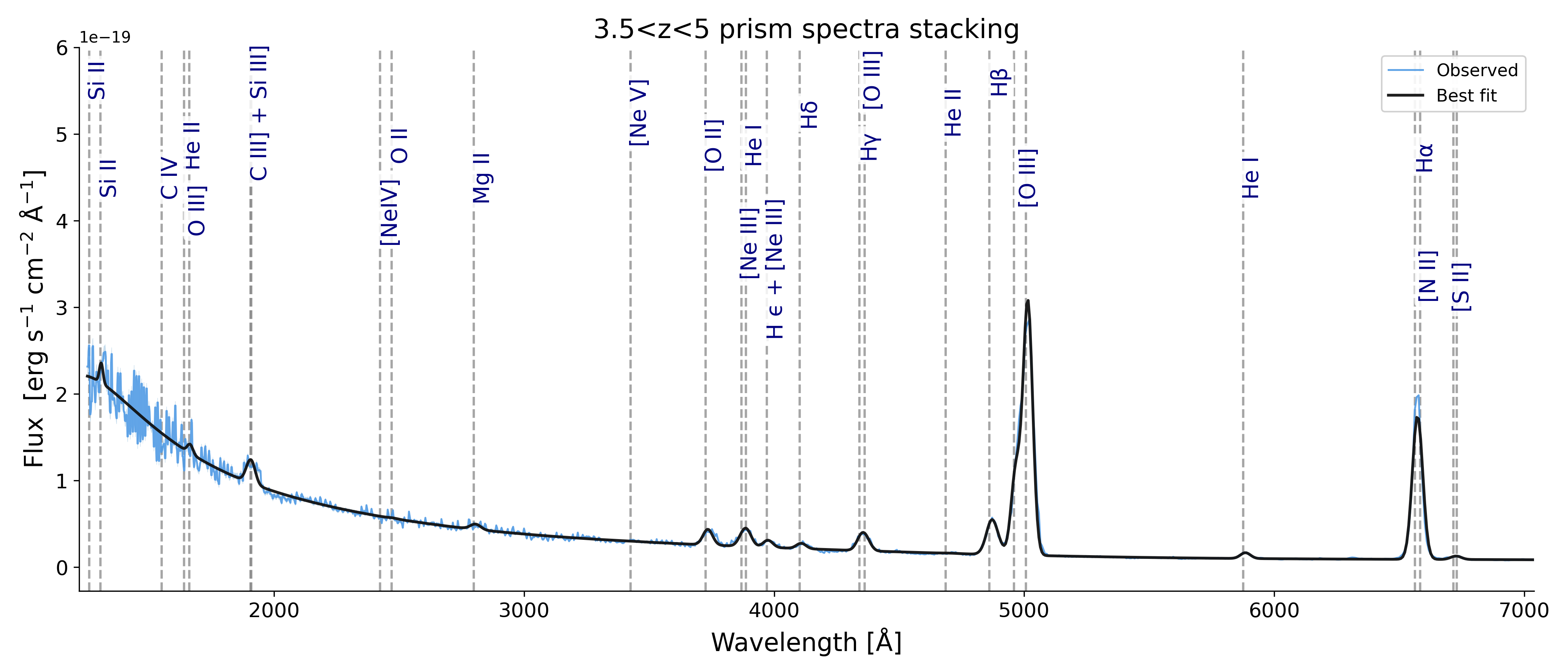}
    \caption{Stack of the rest-frame prism spectra in the  intermediate-redshift bin ($3.5 < z < 5$). Color coding as figures above. For visualization purposes, the normalized stack has been multiplied by the mean [OIII]$\lambda$5007 flux of the 12 contributing objects. This rescaling affects only the plotted figure and not the underlying measurements.}
    \label{fig:z35_z5_prism_stack}
\end{figure}

\begin{figure}[htb]
    \centering
    \vspace{1cm} \includegraphics[width=0.99\textwidth]{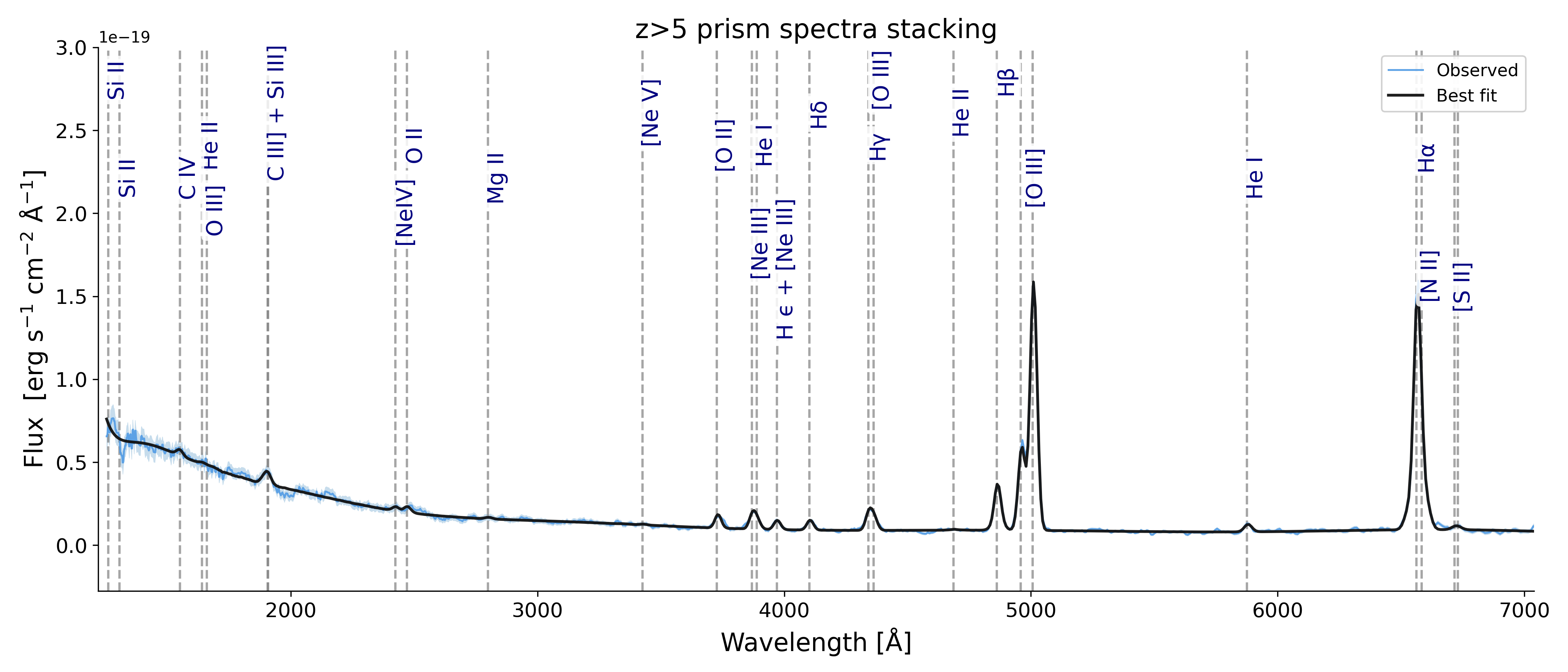}
    \caption{Stack of the rest-frame prism spectra in the high-redshift bin ($z > 5$). Color coding as figures above. For visualization purposes, the normalized stack has been multiplied by the mean [OIII]$\lambda$5007 flux of the 15 contributing objects. This rescaling affects only the plotted figure and not the underlying measurements.}
    \label{fig:z5_prism_stack}
\end{figure}

\end{document}